\documentclass[%
preprint,
superscriptaddress,
nofootinbib,
 amsmath,amssymb,
 aps]{revtex4-1}
\usepackage{graphicx}                               
\usepackage{textcomp}
\usepackage{xcolor}
\usepackage{dcolumn}                                
\usepackage{csquotes}
\usepackage{bm}                                     
\usepackage[version-1-compatibility, load=prefixed, load=abbr, binary-units=true, separate-uncertainty=true]
{siunitx}                                           
\usepackage{booktabs}						 		
\usepackage{multirow}			 			 		
\usepackage{natbib}

\newcommand{\avg}[1]{\langle{#1}\rangle}

\newcommand{\aanda}{Astron. Astrophys.}				
\newcommand{\aph}{Astropart. Phys.} 						
\newcommand{\cpc}{Comput. Phys. Commun.}			
\newcommand{\jgeors}{J. Geophys. Res.-Space} 		
\newcommand{\lrsp}{Living Rev. Sol. Phys.} 				
\newcommand{\nimpra}{Nucl. Instrum. Meth. A.} 		
\newcommand{\physrev}{Phys. Rev.} 						
\newcommand{\solphys}{Sol. Phys.} 							

\begin{document}

\preprint{APS/123-QED}

\title{Measurements of the Time-Dependent Cosmic-Ray Sun Shadow with Seven Years of IceCube Data -- Comparison with the Solar Cycle and Magnetic Field Models}%
\affiliation{III. Physikalisches Institut, RWTH Aachen University, D-52056 Aachen, Germany}

\affiliation{III. Physikalisches Institut, RWTH Aachen University, D-52056 Aachen, Germany}
\affiliation{Department of Physics, University of Adelaide, Adelaide, 5005, Australia}
\affiliation{Dept. of Physics and Astronomy, University of Alaska Anchorage, 3211 Providence Dr., Anchorage, AK 99508, USA}
\affiliation{Dept. of Physics, University of Texas at Arlington, 502 Yates St., Science Hall Rm 108, Box 19059, Arlington, TX 76019, USA}
\affiliation{CTSPS, Clark-Atlanta University, Atlanta, GA 30314, USA}
\affiliation{School of Physics and Center for Relativistic Astrophysics, Georgia Institute of Technology, Atlanta, GA 30332, USA}
\affiliation{Dept. of Physics, Southern University, Baton Rouge, LA 70813, USA}
\affiliation{Dept. of Physics, University of California, Berkeley, CA 94720, USA}
\affiliation{Lawrence Berkeley National Laboratory, Berkeley, CA 94720, USA}
\affiliation{Institut f{\"u}r Physik, Humboldt-Universit{\"a}t zu Berlin, D-12489 Berlin, Germany}
\affiliation{Fakult{\"a}t f{\"u}r Physik {\&} Astronomie, Ruhr-Universit{\"a}t Bochum, D-44780 Bochum, Germany}
\affiliation{Universit{\'e} Libre de Bruxelles, Science Faculty CP230, B-1050 Brussels, Belgium}
\affiliation{Vrije Universiteit Brussel (VUB), Dienst ELEM, B-1050 Brussels, Belgium}
\affiliation{Dept. of Physics, Massachusetts Institute of Technology, Cambridge, MA 02139, USA}
\affiliation{Dept. of Physics and Institute for Global Prominent Research, Chiba University, Chiba 263-8522, Japan}
\affiliation{Department of Physics, Loyola University Chicago, Chicago, IL 60660, USA}
\affiliation{Dept. of Physics and Astronomy, University of Canterbury, Private Bag 4800, Christchurch, New Zealand}
\affiliation{Dept. of Physics, University of Maryland, College Park, MD 20742, USA}
\affiliation{Dept. of Astronomy, Ohio State University, Columbus, OH 43210, USA}
\affiliation{Dept. of Physics and Center for Cosmology and Astro-Particle Physics, Ohio State University, Columbus, OH 43210, USA}
\affiliation{Niels Bohr Institute, University of Copenhagen, DK-2100 Copenhagen, Denmark}
\affiliation{Dept. of Physics, TU Dortmund University, D-44221 Dortmund, Germany}
\affiliation{Dept. of Physics and Astronomy, Michigan State University, East Lansing, MI 48824, USA}
\affiliation{Dept. of Physics, University of Alberta, Edmonton, Alberta, Canada T6G 2E1}
\affiliation{Erlangen Centre for Astroparticle Physics, Friedrich-Alexander-Universit{\"a}t Erlangen-N{\"u}rnberg, D-91058 Erlangen, Germany}
\affiliation{Physik-department, Technische Universit{\"a}t M{\"u}nchen, D-85748 Garching, Germany}
\affiliation{D{\'e}partement de physique nucl{\'e}aire et corpusculaire, Universit{\'e} de Gen{\`e}ve, CH-1211 Gen{\`e}ve, Switzerland}
\affiliation{Dept. of Physics and Astronomy, University of Gent, B-9000 Gent, Belgium}
\affiliation{Dept. of Physics and Astronomy, University of California, Irvine, CA 92697, USA}
\affiliation{Karlsruhe Institute of Technology, Institut f{\"u}r Kernphysik, D-76021 Karlsruhe, Germany}
\affiliation{Dept. of Physics and Astronomy, University of Kansas, Lawrence, KS 66045, USA}
\affiliation{SNOLAB, 1039 Regional Road 24, Creighton Mine 9, Lively, ON, Canada P3Y 1N2}
\affiliation{Department of Physics and Astronomy, UCLA, Los Angeles, CA 90095, USA}
\affiliation{Department of Physics, Mercer University, Macon, GA 31207-0001, USA}
\affiliation{Dept. of Astronomy, University of Wisconsin, Madison, WI 53706, USA}
\affiliation{Dept. of Physics and Wisconsin IceCube Particle Astrophysics Center, University of Wisconsin, Madison, WI 53706, USA}
\affiliation{Institute of Physics, University of Mainz, Staudinger Weg 7, D-55099 Mainz, Germany}
\affiliation{Department of Physics, Marquette University, Milwaukee, WI, 53201, USA}
\affiliation{Institut f{\"u}r Kernphysik, Westf{\"a}lische Wilhelms-Universit{\"a}t M{\"u}nster, D-48149 M{\"u}nster, Germany}
\affiliation{Bartol Research Institute and Dept. of Physics and Astronomy, University of Delaware, Newark, DE 19716, USA}
\affiliation{Dept. of Physics, Yale University, New Haven, CT 06520, USA}
\affiliation{Dept. of Physics, University of Oxford, Parks Road, Oxford OX1 3PU, UK}
\affiliation{Dept. of Physics, Drexel University, 3141 Chestnut Street, Philadelphia, PA 19104, USA}
\affiliation{Physics Department, South Dakota School of Mines and Technology, Rapid City, SD 57701, USA}
\affiliation{Dept. of Physics, University of Wisconsin, River Falls, WI 54022, USA}
\affiliation{Dept. of Physics and Astronomy, University of Rochester, Rochester, NY 14627, USA}
\affiliation{Oskar Klein Centre and Dept. of Physics, Stockholm University, SE-10691 Stockholm, Sweden}
\affiliation{Dept. of Physics and Astronomy, Stony Brook University, Stony Brook, NY 11794-3800, USA}
\affiliation{Dept. of Physics, Sungkyunkwan University, Suwon 16419, Korea}
\affiliation{Institute of Basic Science, Sungkyunkwan University, Suwon 16419, Korea}
\affiliation{Dept. of Physics and Astronomy, University of Alabama, Tuscaloosa, AL 35487, USA}
\affiliation{Dept. of Astronomy and Astrophysics, Pennsylvania State University, University Park, PA 16802, USA}
\affiliation{Dept. of Physics, Pennsylvania State University, University Park, PA 16802, USA}
\affiliation{Dept. of Physics and Astronomy, Uppsala University, Box 516, S-75120 Uppsala, Sweden}
\affiliation{Dept. of Physics, University of Wuppertal, D-42119 Wuppertal, Germany}
\affiliation{DESY, D-15738 Zeuthen, Germany}

\author{M. G. Aartsen}
\affiliation{Dept. of Physics and Astronomy, University of Canterbury, Private Bag 4800, Christchurch, New Zealand}
\author{R. Abbasi}
\affiliation{Department of Physics, Loyola University Chicago, Chicago, IL 60660, USA}
\author{M. Ackermann}
\affiliation{DESY, D-15738 Zeuthen, Germany}
\author{J. Adams}
\affiliation{Dept. of Physics and Astronomy, University of Canterbury, Private Bag 4800, Christchurch, New Zealand}
\author{J. A. Aguilar}
\affiliation{Universit{\'e} Libre de Bruxelles, Science Faculty CP230, B-1050 Brussels, Belgium}
\author{M. Ahlers}
\affiliation{Niels Bohr Institute, University of Copenhagen, DK-2100 Copenhagen, Denmark}
\author{M. Ahrens}
\affiliation{Oskar Klein Centre and Dept. of Physics, Stockholm University, SE-10691 Stockholm, Sweden}
\author{C. Alispach}
\affiliation{D{\'e}partement de physique nucl{\'e}aire et corpusculaire, Universit{\'e} de Gen{\`e}ve, CH-1211 Gen{\`e}ve, Switzerland}
\author{N. M. Amin}
\affiliation{Bartol Research Institute and Dept. of Physics and Astronomy, University of Delaware, Newark, DE 19716, USA}
\author{K. Andeen}
\affiliation{Department of Physics, Marquette University, Milwaukee, WI, 53201, USA}
% ...
\date{\today}

\collaboration{IceCube Collaboration}
\noaffiliation
\date{\today}
\begin{abstract}
Observations of the time-dependent cosmic-ray Sun shadow have been proven as a valuable diagnostic for the assessment of solar magnetic field models.
In this paper, seven years of IceCube data are compared to solar activity and solar magnetic field models.
A quantitative comparison of solar magnetic field models with IceCube data on the event rate level
is performed for the first time. 
Additionally, a first energy-dependent analysis is presented and compared to recent predictions. 
We use seven years of IceCube data for the Moon and the Sun and compare them to simulations on data rate level. The simulations are performed for the geometrical shadow hypothesis for the Moon and the Sun and for a cosmic-ray propagation model governed by the solar magnetic field for the case of the Sun.
We find that a linearly decreasing relationship between Sun shadow strength and solar activity is preferred over a constant relationship at the $6.4\sigma$ level.
We test two commonly used models of the coronal magnetic field, both combined with a Parker spiral, by modeling cosmic-ray propagation in the solar magnetic field.
Both models predict a weakening of the shadow in times of high solar activity as it is also visible in the data. We find tensions with the data on the order of $3\sigma$ for both models, assuming only statistical uncertainties. The magnetic field model \textit{CSSS} fits the data slightly better than the \textit{PFSS} model. This is generally consistent with what is found previously by the Tibet AS-$\gamma$ Experiment, a deviation of the data from the two models is, however, not significant at this point.
Regarding the energy dependence of the Sun shadow, we find indications that the shadowing effect increases with energy during times of high solar activity, in agreement with theoretical predictions.
\end{abstract}
\maketitle
\section{\label{sec:intro}Introduction}
The existence of the cosmic-ray Sun shadow, which commonly refers to cosmic rays being blocked by the Sun, has first been suggested by George W. Clark in 1957 \citep{Clark1957}.
While cosmic rays that propagate close to the Moon are only deflected marginally by the geomagnetic and heliospheric magnetic fields, cosmic rays that traverse the coronal solar magnetic field can be deflected strongly and irregularly.

Hence, the cosmic-ray Moon shadow essentially blocks cosmic rays from a well-known solid angle and can be used as a direction and resolution standard.
Several experiments have exploited this feature to study their angular resolution, absolute pointing and absolute energy scale\footnote{The latter is done by using the energy-dependent shift of the position of the shadow due to the Earth's magnetic field.}, e.g.\ Tibet AS-$\gamma$, MILAGRO, ARGO-YBJ and IceCube \citep{Tibet1993Moon, Milagro2001, Argo2011Moon, IceCube2014}.

The cosmic-ray Sun shadow, on the other hand, contains the footprint of the solar magnetic field in the form of those cosmic rays reaching Earth that come from directions close to the Sun.
In 2013, the Tibet AS-$\gamma$ Collaboration compared coronal magnetic field models using the cosmic-ray Sun shadow at a median primary cosmic-ray energy of $\sim \SI{10}{\TeV}$ \citep{Tibet2013}. 
Later, they also studied the influence of solar coronal mass ejections (CMEs) on the cosmic-ray Sun shadow at energies of $\sim \SI{3}{\TeV}$ \citep{Tibet2018} and concluded that Earth-directed CMEs (ECMEs) affect the cosmic-ray Sun shadow at these energies.
The influence of the solar magnetic field on the cosmic-ray Sun shadow has been studied by several other experiments, like Milagro, ARGO-YBJ, and HAWC, as well \citep{Milagro2003, Argo2011, Hawc2015}.
Such efforts are especially important since no in-situ measurements of the solar magnetic field closer than to about \SI{0.29}{\astronomicalunit} ($\approx 62$ solar radii) distance from the Sun (\textit{Helios} spacecraft \citep{Helios1981}) existed until very recently. Even the \textit{Parker Solar Probe}, which will eventually approach the Sun up to $\sim 8.9$ solar radii in 2024 \cite{Parker2019}, has as yet not been closer than 27.8 solar radii, i.e.\ about 0.13 AU.

In this paper, we investigate  a possible time dependence  of the cosmic-ray Sun shadow using seven years of IceCube data. The measurements are based on atmospheric muons detected with IceCube, induced by cosmic rays entering the Earth's atmosphere. For the first time, we use IceCube data for a comparison of measurements with the expected shadow for solar activity with different solar magnetic field models. We calculate the median energy of the primary cosmic rays in our sample to $50-60$~TeV, depending on the cosmic-ray flux model that is used to derive cosmic-ray energies from the measured muons. This is therefore the highest energy measurement of the Sun shadow so far, as compared to the median primary cosmic-ray energies of the earlier measurements that lie around $1-10$~TeV. 
Additionally, the energy-dependence of the Sun shadow is investigated, i.e.\ two samples with respective median energies of $40$~TeV and $100$~TeV are produced and qualitatively compared to a recent prediction, in which the energy-dependence of the Sun shadow in a low-activity solar magnetic field is shown to differ significantly from that in a high-activity solar magnetic field \citep{beckertjus2020}.
The data we use for studying the cosmic-ray Sun shadow in IceCube comprise the time from late 2010 till early 2017 and cover large parts of Solar Cycle 24. This cycle is defined for the time interval from late 2008 to some time around late 2019 to early 2020, wherein the exact location of the minimum that defines the end of the cycle is not clear at this point.
This work follows an earlier study which reported on the detection of a temporal variation of the cosmic-ray Sun shadow measured with IceCube and found a correlation with solar activity to be likely \citep{MoonSunApj2019}.
\section{\label{sec:icecube}The IceCube Neutrino Observatory}

The IceCube Neutrino Observatory is a detection array deployed in the Antarctic ice near the geographic South Pole and comprises a volume of about \SI{1}{\cubic\kilo\meter} instrumented with 5160 Digital Optical Modules (DOMs) on 86 strings \citep{icecube_detector2017}. 
IceCube is located at a depth between \SI{1450}{\meter} and \SI{2450}{\meter} and detects relativistic secondary particles induced by astrophysical neutrinos, gamma rays, and cosmic rays.
The detector was built at the South Pole between 2005 and 2010 and exploits the clear Antarctic ice as its detection medium for Cherenkov radiation of charged particles traversing it. 

\begin{figure}[htbp]
    \centering
    \includegraphics[width=\linewidth]{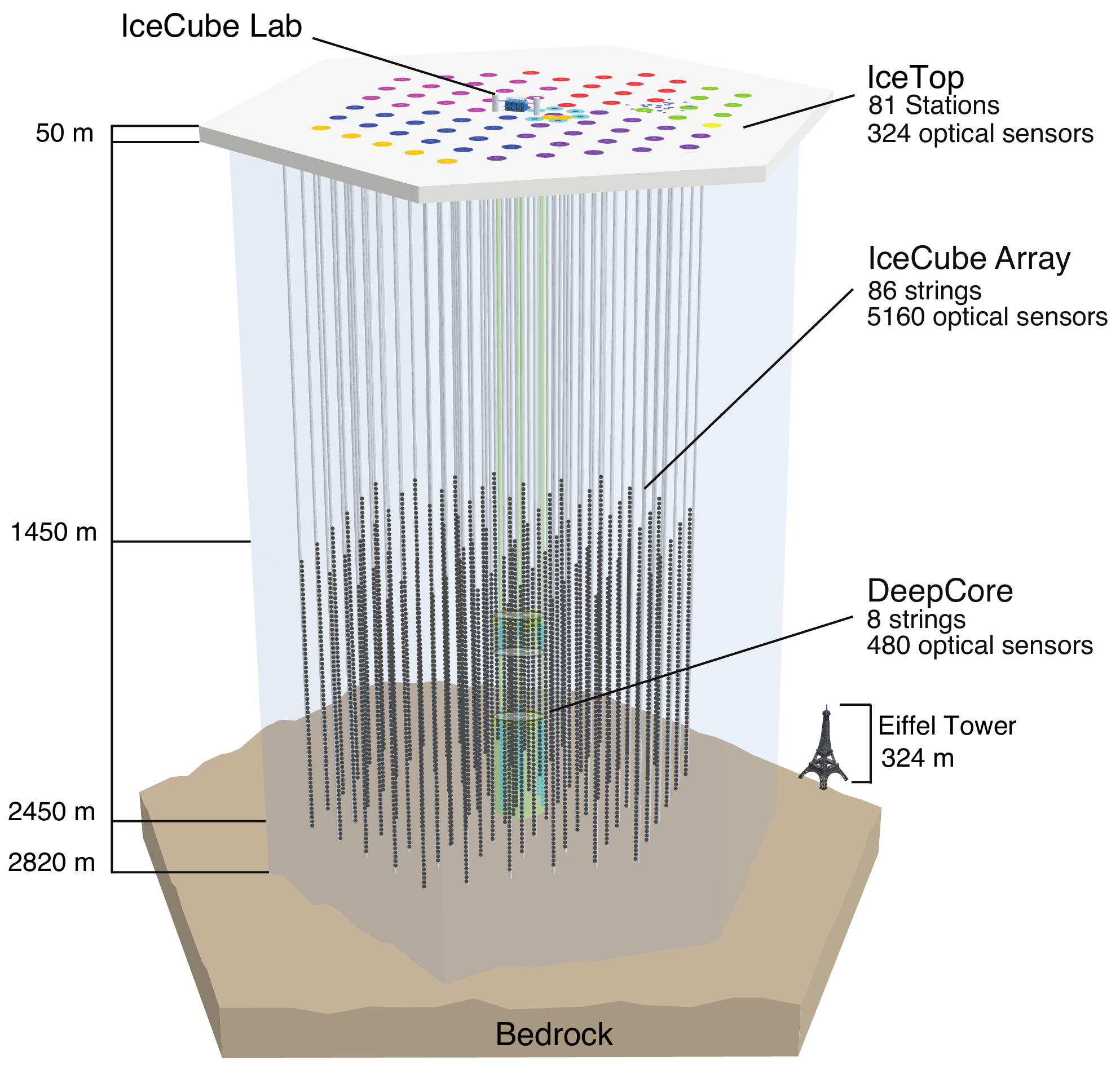}
    \caption{The IceCube Neutrino Observatory.
    }
    \label{fig:icecube}
\end{figure}

A sketch of the IceCube Neutrino Observatory including its sub-array DeepCore \citep{deepcore2009}, which aims to improve the sensitivity to lower-energy neutrinos, can be seen in Figure \ref{fig:icecube}. Data from DeepCore have not been used in this analysis.
For neutrinos, IceCube's main array has an energy threshold of about \SI{100}{\GeV}. In this paper, we use atmospheric muons, which are a background to the neutrino searches. In this sample, the energies of the primary cosmic rays inducing these atmospheric muon events are typically $\gtrsim 1$~TeV.
{}
%
\section{\label{sec:selection}Data sample}

In IceCube, high-energy muons are observed. As these are predominantly produced by cosmic-ray air showers, they trace the direction of the primary particles, with the angular uncertainty being dominated by the uncertainty of the light propagation in the ice and limited by the kinematic angle between primary and secondary particle. We thus measure an event rate of high-energy muons in IceCube.
The event rate increases with increasing elevation because of the decreasing amount of ice overburden that cosmic-ray induced atmospheric muons have to cross to reach the detector. In this section, we describe the details of our data sample.

\subsection{\label{sec:selection_pole}Moon and Sun as seen from the South Pole}
This paper uses data from IceCube's 79-string configuration (IC79), which was available in the 2010/2011 season and from the final 86-string configuration (IC86), which was available from the the 2011/2012 season and onwards. We use the high-energy atmospheric muons that pass through the detector for our analysis, as they are direct tracers of the primary particles. 
The strength of the cosmic-ray shadow of the Moon and Sun is determined by the number of cosmic rays that are blocked. 
Without additional forces this number results from the solid angle that is spanned by the Moon and the Sun as seen from Earth, i.e.\ their angular radii. 
The minimum and maximum angular radii of Moon and Sun as seen from the South Pole amount to \SIrange{0.245}{0.279}{\degree} for the Moon and \SIrange{0.262}{0.271}{\degree} for the Sun. 
Notably, both objects have an angular diameter of $\sim \SI{0.5}{\degree}$, which makes a comparison relatively straightforward.

\begin{figure}[htbp]
    \centering
    \includegraphics[width=\linewidth]{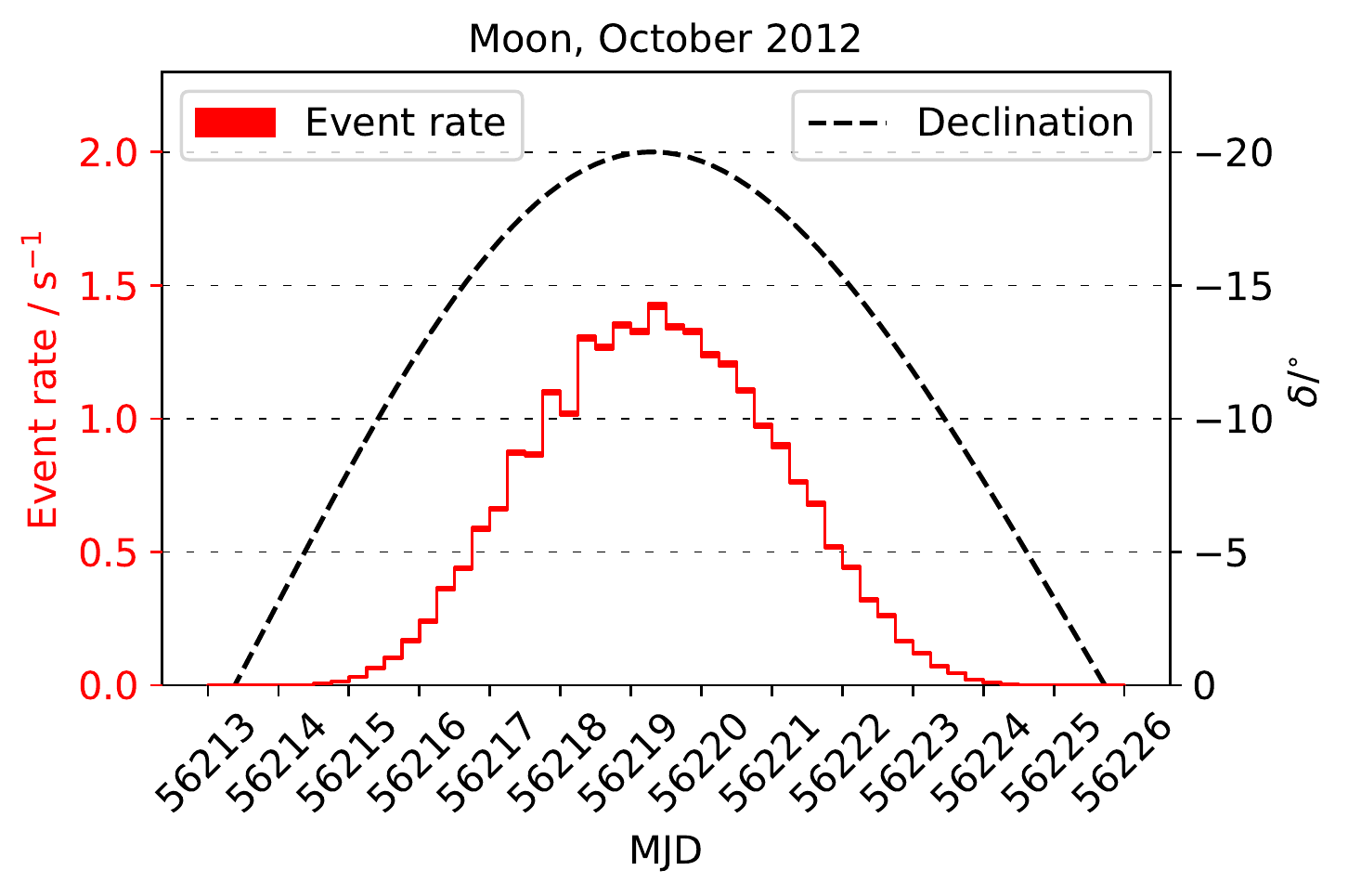}
    \caption{Comparison of Moon declination $\delta$ and final Moon shadow sample event rate. At the South Pole, the elevation of an object equals $-\delta$}
    \label{fig:MoonDec}
\end{figure}

The maximum elevation of the Moon at the South Pole varies between \ang{18.3} and \ang{28.6} due to its orbital inclination and Earth's axial tilt. 
While the Earth axial tilt changes only very slowly, the Moon's orbital inclination varies with a nodal period of about 18.6 years.
For the Sun's elevation, on the other hand, only the Earth axial tilt is relevant for its maximum elevation, which amounts to about \ang{23.4} each year. 
The Sun rises and sets only once per year at the South Pole, resulting in one continuous observation period approximately from November through February each year. The Moon instead rises and sets approximately every 27 days, which leads to about 13 separate observation periods. 
An example one such period is shown in Figure \ref{fig:MoonDec}. 
{}
%
\subsection{\label{ssec:selection_online}Data selection and quality cuts}
At the South Pole, the direction of each muon event is reconstructed using a fast maximum-likelihood method. 
Since atmospheric muons detected in IceCube are ultra-relativistic, the opening angle between muon direction and primary cosmic-ray direction, which is on the order of \ang{0.1} for multi-TeV muons \citep{IceCube2014}, is part of the directional uncertainty between the actual primary cosmic-ray direction and the reconstructed event direction.
 Data are only taken when Moon and Sun are above the horizon (at least \ang{15} above the horizon for data between May 2010 and May 2012, as in these early years, directional reconstruction near the horizon was not good enough). 
These so-called Moon and Sun filters are implemented at the South Pole and are necessary in order to reduce the data to a manageable amount for satellite data transmission to the Northern hemisphere. 

For further event reconstruction, a $\pm \ang{10}$ zenith band ($\pm \ang{180}$ in azimuth) around the known position of Moon/Sun in the sky is considered. While this full azimuth band is available at the low-level of the analysis, the 6 off-source regions that we chose only make use of parts of this band, in total $\ang{54} \times \ang{6}$. The reason not to include more off-source regions is that it would increase the processing (and final data file sizes), while not substantially reducing the statistical uncertainty.

The events are then selected with the requirement to hit at least 8 DOMs on three different strings.

After the first selection at the South Pole, the data are transferred North and more sophisticated reconstruction algorithms are applied to the following parameters:
Besides the multi-photo-electron (MPE) fit \citep{Amanda2004}, which accounts for the total number of Cherenkov photons collected by each DOM, this includes a paraboloid fit to the likelihood profile of the directional coordinates \citep{Neunhoeffer2006}.

To ensure that only well-reconstructed events are used for the final data analysis, two quality cuts based on these reconstructions are applied:
\begin{enumerate}
    \item The reduced log-likelihood (rlogl), which represents the goodness-of-fit of the MPE reconstruction, is required to satisfy $\mathrm{rlogl} < 8.1$, see \citep{Neunhoeffer2006}.
    \item The angular uncertainty $\sigma$, which is derived from the paraboloid fit to the likelihood profile\footnote{The likelihood profile is defined as the entity of likelihood values as a function of the directional coordinates, see \cite{Neunhoeffer2006} for details.}, is required to satisfy $\sigma <0.71^{\circ}$.
\end{enumerate}{}
Both quality cuts were determined with the goal to maximize the statistical significance of the shadows in \cite{Bos2017} and have been used in previous studies \cite{MoonSunApj2019, Tenholt2019, Tenholt2020}.

\section{\label{sec:analysis}Data analysis}

\subsection{\label{sssec:analysis_method_coord}Relative coordinates}

The direction of each muon in the sample is  compared to the known position of Moon/Sun in the sky and relative local coordinates are calculated. The individual reconstructed declination of each muon event is defined as $\delta_{\mu}$.

The position is given in equatorial coordinates, with the quasi-Cartesian values of the coordinates relative to the center of Moon/Sun given as $x=\Delta \alpha \cos\delta_{\mu}$ and $y=\Delta \delta$, where $\delta$ is the declination and $\alpha$ the right ascension. The sign of the relative difference is defined as  $\alpha/\delta_{\mu}$ - $\alpha/\delta_{\rm Moon/Sun}$.
%
\subsection{\label{sssec:analysis_method_onoff}On- and off-source regions}

\begin{figure}[htbp]
    \centering
    \includegraphics[width=\linewidth]{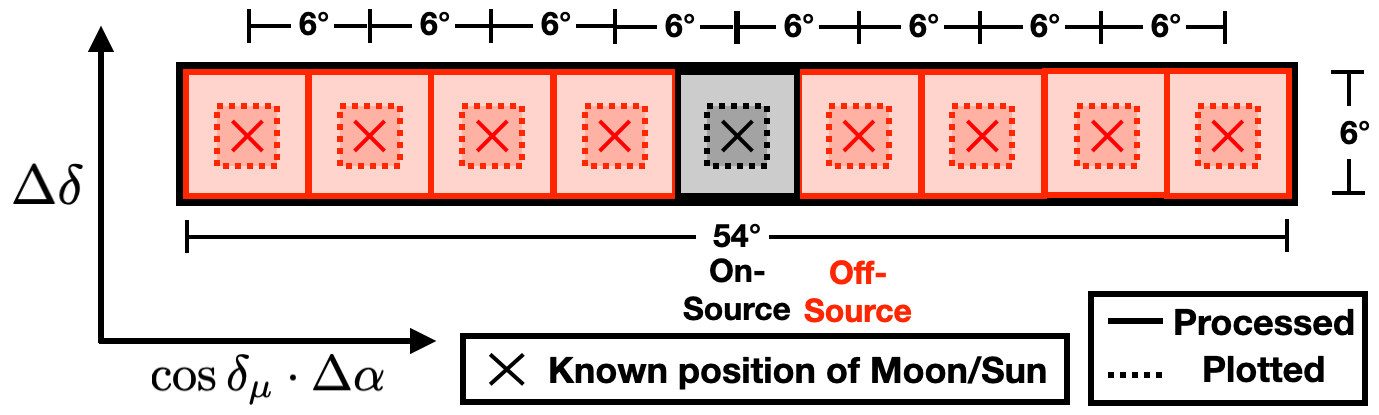}
    \caption{On- and off-source regions used for the data analysis. The black \enquote{x} marks the zero point of the relative coordinates, i.e. the center of Moon/Sun.}
    \label{fig:on_off}
\end{figure}

Based on the calculated quasi-Cartesian relative coordinates $x$ and $y$, one on-source region and eight off-source regions are defined as shown in Figure~\ref{fig:on_off}.
Each region has an angular extent of $\ang{6} \times \ang{6}$, resulting in a total angular area of $\ang{54} \times \ang{6}$ for the nine regions. 

In order to account for the spherical distortion, we keep the corrected relative right ascension $\Delta \alpha\cos\delta_{\mu}$ of the entire analyzed region constant at \ang{54} rather than the uncorrected relative right ascension $\Delta \alpha$.

{}
%

\subsection{\label{sssec:analysis_final_sample}Event numbers and average declination}
The number of events contained in the $\ang{54} \times \ang{6}$ window described in Section \ref{sssec:analysis_method_onoff} and their average declination are given in Table \ref{tab:moon_sun_parameters}. 

It can be seen that the number of events varies between 3.8 and 7.9 million events for the Moon shadow sample, while it amounts to 13.1 to 13.3 million events for the Sun shadow sample except for IC79, which contains about 9 million events.

The average declination, on the other hand, varies between \ang{-16.7} and \ang{-22.1} for the Moon shadow sample and amounts to \ang{-21.8} for the Sun shadow sample except for IC79, where it is slightly smaller with \ang{-22.1}. 
These values are used later for modelling the expected relative deficit due to the lunar and solar disk as described in detail in Section \ref{sssec:analysis_sim_gen}.

\begin{table}[htbp]
\centering
\caption[Moon/Sun data sample parameters]{Number of events and average declination of the data sample for each year.}
\label{tab:moon_sun_parameters} 
\begin{tabular}{@{}lrrrr@{}}
\toprule
 & \multicolumn{2}{c}{Moon} & \multicolumn{2}{c}{Sun} \\ 
Year & $N / \num{e6}$ & $-\avg{\delta} / \si{\degree}$ & $N / \num{e6}$ & $-\avg{\delta} / \si{\degree}$ \\ 
\midrule
IC79    & \num{7.9} & \num{22.1} & \num{9.0}   & \num{22.1} \\
IC86-1 & \num{7.7} & \num{20.7} & \num{13.1} & \num{21.8} \\
IC86-2 & \num{6.4} & \num{18.9} & \num{13.1} & \num{21.8} \\
IC86-3 & \num{4.5} & \num{17.7} & \num{13.2} & \num{21.8} \\
IC86-4 & \num{3.8} & \num{16.9} & \num{13.2} & \num{21.8} \\
IC86-5 & \num{4.1} & \num{16.7} & \num{13.3} & \num{21.8} \\
IC86-6 & \num{5.1} & \num{17.2} & \num{13.3} & \num{21.8} \\
\bottomrule
\end{tabular}
\end{table}
%

\subsection{\label{sssec:analysis_method_maps}2D maps and smoothing}

After defining on- and off-source regions as described in Section \ref{sssec:analysis_method_onoff}, the off-source regions are shifted with respect to the on-source region such that they are centered at $x = \ang{0}$ (displayed as the black $\times$ in Figure \ref{fig:on_off}), becoming directly comparable to the on-source region, which is centered at $x=\ang{0}$ by definition.

Then, two two-dimensional binned histograms containing the number of events are defined: the first encloses the on-source region, the second represents the average of the eight off-source regions.
Both histograms cover $\ang{6} \times \ang{6}$ in $x$ and $y$ and consist of $60 \times 60$ bins $(i,j)$ ($i=1\ldots 60$, $j=1\ldots 60$), wherein each bin has a size of $\ang{0.1} \times \ang{0.1}$.

Then, the relative deficit due to the shadowing of Moon/Sun in each bin $(i,j)$ is calculated using the number of events $N_{\mathrm{on}}^{i,j}$ in bin $(i,j)$ in the on-source region and the average number of events $\avg{N_{\mathrm{off}}}^{i,j}$ in the eight off-source regions:
\begin{align}
\left(\frac{\Delta N}{\avg{N_{\mathrm{off}}}}\right)^{i,j} = \frac{N_{\mathrm{on}}^{i,j} - \avg{N_{\mathrm{off}}}^{i,j}}{\avg{N_{\mathrm{off}}}^{i,j}} \,.
\label{eq:shadows_rd}
\end{align}
The average number of events $\avg{N_{\mathrm{off}}}^{i,j}$ in a bin $(i,j)$ located in the off-source regions is calculated by averaging over the off-source regions:
\begin{align}
\avg{N_{\mathrm{off}}}^{i,j}=\frac{1}{8} \sum_{n=1}^{8}\left(N_{\mathrm{off}}^{i,j}\right)_{n} \,,
\end{align}{}
where $\left(N_{\mathrm{off}}^{i,j}\right)_{n}$ is the number of off-source events in the $n$th off-source region. 
The result is a two-dimensional map of the relative deficit due to the Moon/Sun shadow.
In order to better suppress statistical fluctuations, the two-dimensional relative deficit map is smoothed  with a box-car smoothing algorithm that replaces the relative deficit in each bin with the average of all bins within \ang{0.7} around its center. It is described in Section \ref{sec:results}.
The chosen smoothing radius corresponds approximately to the median angular resolution of the final data sample.

To guide the eye, and for the numerical analysis presented in the next section, the center of gravity of the shadow is determined and plotted as well\footnote{It should be noted that for the Sun shadow, the center of gravity is not necessarily expected to align with the center of the solar disk due to the influence of the solar magnetic field.} (see Fig.\ \ref{fig:shadows_moon_2d} and Fig.\ \ref{fig:shadows_sun_2d}). 
It is determined by averaging over the  positions of all bins with a relative deficit of \SI{3}{\percent} or more after smoothing.
As typical statistical uncertainties after the smoothing amount to about \SI{0.6}{\percent}, this threshold defines bins that show a statistically significant deficit of events. 
\subsection{\label{sssec:analysis_method_num}Numerical analysis}
In order to quantify the deficit of cosmic-ray induced muon events due to the shadowing of the Moon and Sun, the relative deficit of events in a \ang{1}-circle around the center of gravity (cf. previous section) is computed.
Choosing a reasonable search radius is a trade-off between a smaller statistical error on the one hand and more off-source background contamination (washing out the deficit due to the Moon and Sun) on the other hand. 
Within \ang{1} the cumulative point spread function contains about \SI{70}{\percent} of events while the off-source background contamination is still relatively small. 
The statistical uncertainty of the relative deficit is computed using error propagation as
\begin{align}
\sigma_{\mathrm{RD}} = \frac{N_{\mathrm{on}}}{\avg{N_{\mathrm{off}}}} \sqrt{\frac{1}{N_{\mathrm{on}}} + \frac{1}{s \cdot \avg{N_{\mathrm{off}}}}} \,,
\label{eq:shadows_rd_err}
\end{align}
with the number of off-source regions $s=8$.

In addition to the relative deficit, the significance of the shadowing effect is calculated using a standard formula developed by Li and Ma \cite{LiMa1983}, whereby a \ang{0.7}-circle around the center of gravity is chosen as search area. 
The selected search radius maximizes the statistical significance for very large numbers of background events and for an angular resolution typical for atmospheric muon events (cf. \cite{Bos2017,Tenholt2020} for details).
\section{\label{ssec:analysis_sim}Simulations}
\subsection{\label{sssec:analysis_sim_mod}Models}
The simulations used for characterizing the cosmic-ray induced atmospheric muon flux are based on \texttt{CORSIKA} \cite{Heck1998}. More specifically, two \texttt{CORSIKA}-generated simulation sets are used, covering primary energies from $600$~GeV to $100$~EeV
and containing $^1$H, $^4$He, $^{14}$N, $^{27}$Al, and $^{56}$Fe nuclei.
Hadronic interactions are simulated with Sibyll 2.1 \cite{Ahn2009} and the MSIS-E-90 atmospheric profile \cite{Nasa2019}.
Lepton propagation in ice is carried out using the lepton propagation tool \texttt{PROPOSAL} \cite{Proposal2013}.
Light emission and propagation is handled using \texttt{GEANT4} \cite{Geant2003} and the IceCube-internal software package \texttt{CLSim} that has been developed based on the \texttt{Photonics} code \cite{Photonics2007}.
The Antarctic ice in which IceCube is embedded is modeled using the \texttt{SPICE Lea} model \cite{SpiceMie2013, SpiceLea2013}.
The detector response is simulated based on internal software.
After simulating atmospheric muon events using the models described above, each event is weighted based on a model of the primary cosmic-ray flux: 
 The weight $w$ of an event induced by a
primary cosmic ray with energy $E$, mass number $A$, and atomic number $Z$ is determined
as the ratio of the cosmic-ray flux $\Phi_{\rm model}$ according to a chosen model and the simulated
cosmic-ray flux $\Phi_{\rm sim}$:
\begin{equation}
w(E,\,A,\,Z)=\frac{\Phi_{\rm model}(E, A, Z)}{\Phi_{\rm sim}(E, A, Z)}\,.
\end{equation}
Here, the model by Gaisser with an extragalactic component presented in \cite{Gaisser2012} and based on the Hillas approach, thus called \texttt{HGm} model hereafter, is used.
\subsection{\label{ssec:selection_char}Sample characteristics}
\begin{figure}[htbp]
    \centering
    \includegraphics[width=\linewidth]{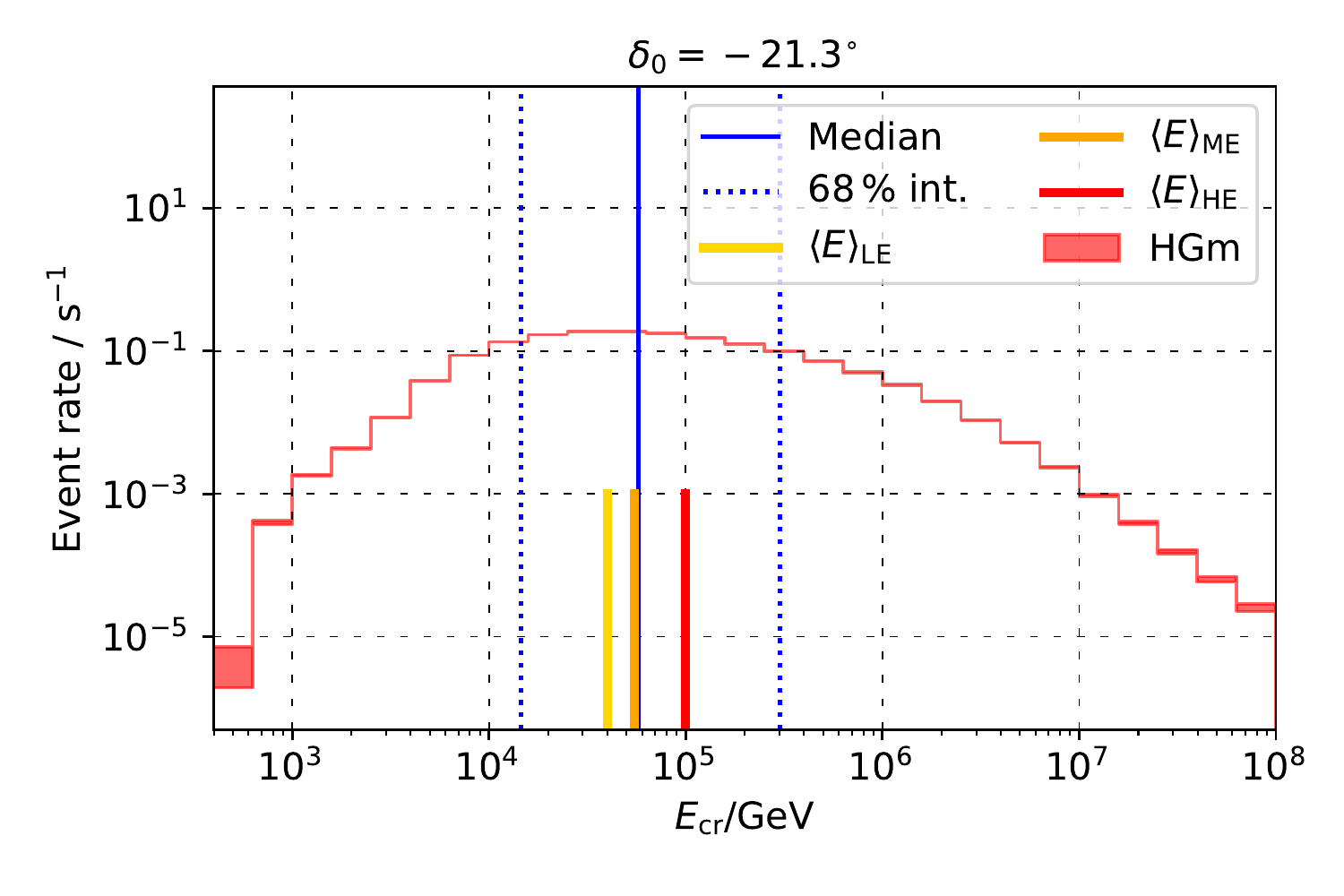}
    \caption{Energy distribution in a $\pm \ang{3}$ declination band around the base declination $\delta_0 = \ang{-21.3}$. The thick yellow, orange, and red lines indicate the median primary cosmic-ray energies studied in Section \ref{ssec:analysis_energy}.}
    \label{fig:EnergyDist}
\end{figure}

\begin{figure}[htbp]
    \centering
    \includegraphics[width=\linewidth]{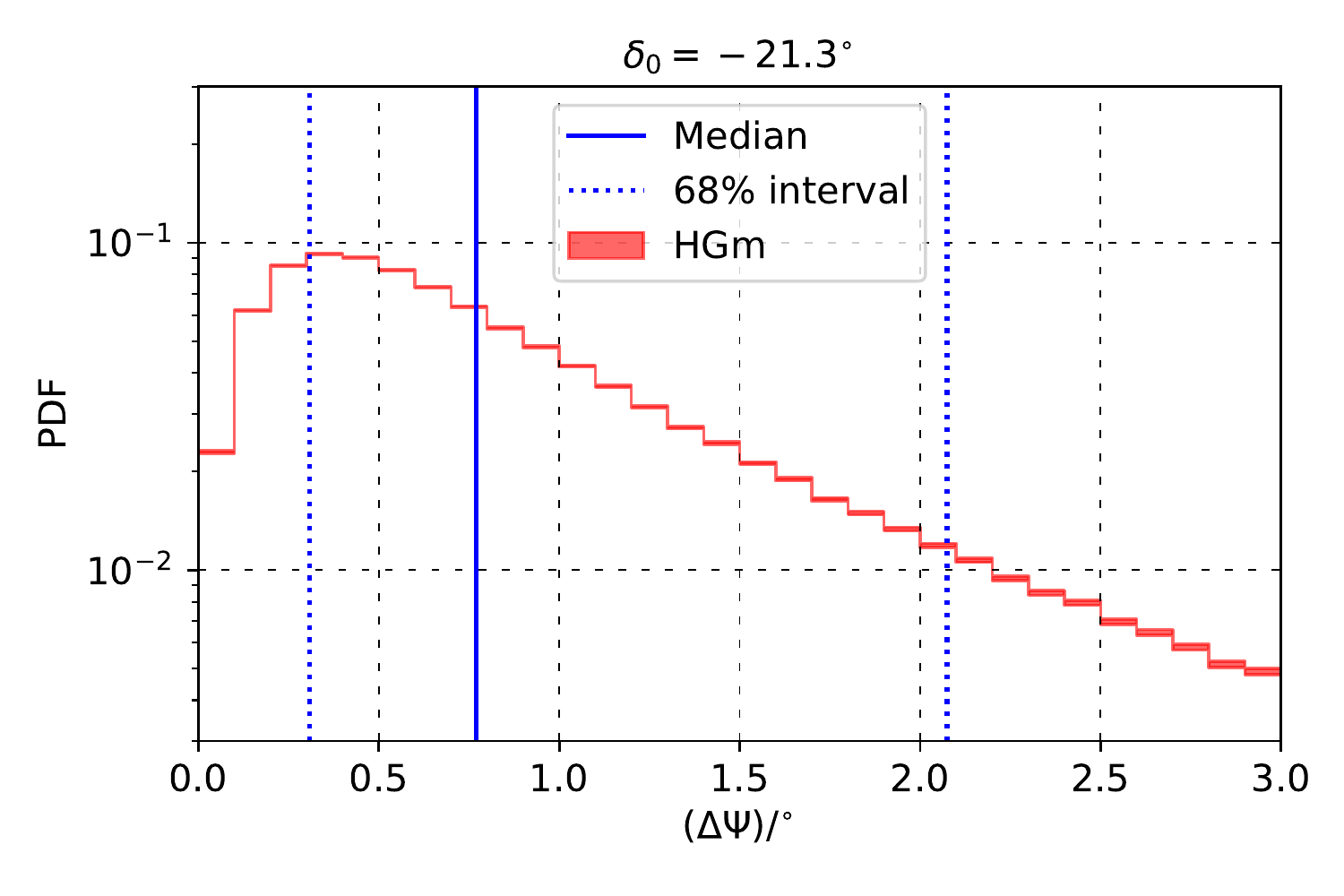}
    \caption{Angular error distribution in a $\pm \ang{3}$ declination band around the base declination $\delta_0 = \ang{-21.3}$.}
    \label{fig:AngularError}
\end{figure}

Based on the models described in the previous section and the data analysis presented in Section~\ref{sec:analysis}, the (simulated) data sample is characterized with respect to the energy distribution of primary cosmic rays (Figure \ref{fig:EnergyDist}) and the probability density function (PDF) of the opening angle $\Delta \theta$ between reconstructed muon direction and actual cosmic-ray direction (Figure \ref{fig:AngularError}).
Both figures show the distribution of the final simulation sample after the same steps as described in Section~\ref{sec:selection}.
The base declination $\delta_0$ is chosen such that the $\pm \ang{3}$ declination band around it has the same average declination as the final Sun shadow data sample (cf. \cite{Tenholt2020} for more details).
{}
The simulation sample contains events with primary energies between $\sim \SI{1}{\TeV}$ and $\sim \SI{100}{\peta\eV}$.
The median energy amounts to about \SI{60}{\TeV} and the \SI{68}{\percent} interval is between \SI{14.6}{\TeV} and \SI{302}{\TeV}.
{}
The median angular error amounts to $\ang{0.77}$, the \SI{68}{\percent} interval covers values between \ang{0.31} and \ang{2.1}. 
\subsection{\label{sssec:analysis_sim_gen}Simulating the shadows}
The shadowing of cosmic rays due to the Moon and Sun is modeled by modifying the primary cosmic-ray weight $w$ of each event.
Using the probability $p$ of each primary cosmic ray to pass through interplanetary space without hitting the Moon or the Sun, the modified weight $w'$ is calculated as 
\begin{align}
w' = p \cdot w \,.
\label{eq:sim_modified_weight}
\end{align}

In the simplest model, the Moon and Sun are treated as non-magnetic, totally absorbing spheres in space which block those cosmic rays that come from directions within the respective lunar and solar disk as seen from Earth.
In this model, $p$ is a step function only depending on the space angle $\Delta \theta$ between the cosmic-ray direction and the center of the Moon/Sun.

Although cosmic rays are deflected by the geomagnetic field, the net shadowing effect remains largely unaffected, besides a small shift of the shadow that is significantly smaller than the resolution of the detector is expected. 
Moreover, by applying the center-of-gravity correction presented in Section \ref{sssec:analysis_method_num} before calculating the relative deficit, such a shift is accounted for in the analysis method. 

In order to simulate the expected relative deficit due to the lunar and solar disk, two key parameters are taken into account: the average declination $\avg{\delta}$ of each data sample given in Table \ref{tab:moon_sun_parameters} and the weighted average of the apparent radius $\langle R_{\rm app}^{\rm Moon/\odot}\rangle$ of Moon and Sun.
While the average declination determines typical energies and the median angular resolution (cf. \cite{Tenholt2020}), the apparent radius determines, in simple words, how large lunar and solar disk have to be modeled.

For calculating $\langle R_{\rm app}^{\rm Moon/\odot}\rangle$, the number of events for each Modified Julian Date (MJD), $N_{\rm MJD}$, is determined together with the apparent radius in the sky for each individual MJD of the sample, $R_{\rm app}^{\rm MJD}$. 
The latter is calculated by obtaining the Earth-Moon distance for that specific MJD and
applying trigonometry. Using these pieces of information, the weighted apparent radius becomes 
\begin{equation}
    \langle R_{\rm app}^{\rm Moon/\odot}\rangle=\frac{\sum_{\rm MJD}N_{\rm MJD}\cdot R_{\rm app}^{\rm MJD}}{\sum_{\rm MJD}N_{\rm MJD}}\,.
\end{equation}
With this procedure, the weighted average of the apparent radius of the Moon in the sky is shown to vary between the $0.251^{\circ}$ (IC86-6) and $0.274^{\circ}$ (IC86-2). 

For the Sun, the range of its apparent size varies, in general, between $0.262^{\circ}$ and $0.271^{\circ}$ over the year. 
For the time from November through February studied in this paper, however, it amounts to $0.271^{\circ}$ for each year.
As a result of neglecting the sub-percent variation of this value, the expectation for the geometrical shadowing effect of the solar disk is the same in each year.

The expected relative deficit due to the shadow induced by the disk-model described above is referred to as lunar disk and solar disk in Figures \ref{fig:MoonDisk} and \ref{fig:SunDisk}.

For simulating the Sun shadow including the effect of the solar magnetic field, the picture is much more complex.
Besides the geomagnetic field, also the heliospheric magnetic field and especially the solar coronal magnetic field deflect cosmic rays. 
While the heliospheric magnetic field, like the geomagnetic field, is comparably regular, the coronal magnetic field can become highly irregular. This increased level of magnetic small-scale variability enhances the interactions of cosmic rays with the magnetic field, thus changing the net shadowing effect of the Sun. A first quantification of how the shadowing effect is changed has been discussed in \citep{beckertjus2020}.
It is thus necessary to actually simulate cosmic-ray propagation in the heliospheric and coronal magnetic field. In such a simulation,  the passing probability $p$ of each primary cosmic ray can be determined. 
It can 
be calculated as the number of cases $n_{\rm pass}$ in which the cosmic-ray particle traverses the solar corona
without hitting the photosphere, divided by the total number of trials $n_{\rm total}$,
\begin{equation}
p= \frac{n_{\rm pass}}{n_{\rm total}}  \,.
\end{equation}
This probability can be calculated by propagating cosmic rays in the magnetic field of the Sun. Here, we use a backtracking approach for computing time-efficient simulations and we perform the propagation in two different magnetic field models. The details of these parts of the simulation are described in the subsequent Sections \ref{sssec:analysis_sim_back} and \ref{ssec:analysis_sim_models}.

\subsection{\label{sssec:analysis_sim_back}Particle propagation in the solar magnetic field}

In order to produce simulations of the cosmic-ray shadow at Earth, the first step is to numerically propagate the particles through the magnetic field of the Sun. We use the test-particle approach, which implies that the magnetic field configuration is not changed by the particle current. This is a reasonable assumption for such high-energetic particles, whose coronal crossing time of a few minutes is much smaller than the timescales of solar magnetic variability, and thus allows us to keep the magnetic field configuration constant for the simulation of one particle trajectory.

The particles are propagated according to the equation of motion
\begin{equation}
    \frac{d\vec{p}}{dt}=q\,\left(\vec{v}\times \vec{B}\right)
\end{equation}
following the approach in \cite{beckertjus2020}.
The simulations are performed for different magnetic field configurations, corresponding to the solar magnetic field at different times. In particular, simulations for two magnetic field models are performed as described in the following subsection. These fields are provided for each month during the measurement and are kept constant for that month. The details of this procedure are described in \cite{beckertjus2020}.

Propagating particles forward is very inefficient, as it cannot be defined beforehand which of these particles actually hit Earth and which do not. Thus, in order to produce a computing-time efficient simulation,  only those cosmic rays that eventually induce an atmospheric muon event in the final simulation sample are propagated. 
This is achieved by using a backtracking method, which takes the known primaries of the final simulation sample, changes all particles into their anti-particles and, at the same time, inverts their momentum vector.  This means that in the simulations, we start anti-particles at Earth, propagate them around the Sun and detect the resulting projected shadow behind the Sun. Changing the charge and the direction at the same time delivers the same result as the propagation of particles along the inverted path. The back-tracking method is therefore well-suited to reduce computational time while still providing a proper picture of the propagation in the magnetic field.
\subsection{\label{ssec:analysis_sim_models}Solar magnetic field models}
As mentioned before, the solar magnetic field consists of two components: the coronal magnetic field and the heliospheric magnetic field.
While the coronal magnetic field is  modeled using (a) a potential-field model and (b) a magnetohydrostatic model, the heliospheric magnetic field is modeled using a Parker spiral approach \citep{Parker1958}, with a semi-linear approximation of the radial solar wind velocity profile.  
\subsubsection{PFSS model}
The potential-field source-surface (PFSS) model \citep{Schatten1969, Altschuler1969} assumes the solar corona to be force- and current-free, i.e. $\vec{j}=\vec{0}$, with the current density $\vec{j}$.
Neglecting the displacement current, $\vec{j}$ can be related to the curl of the magnetic field $\vec{B}$ as 
\begin{align}
\vec{j} = \frac{1}{\mu_{0}} \left(\vec{\nabla} \times \vec{B}\right)\,.
\end{align}{}
For a current-free corona the magnetic field must hence be curl-free, $\vec{\nabla} \times \vec{B} = \vec{0}$, which means that it can be expressed as the gradient of a scalar potential $\Phi$, \mbox{$\vec{B} = - \nabla \Phi$}.
With $\vec{\nabla} \cdot \vec{B} = 0$, this yields the Laplace equation
\begin{align}
\vec{\nabla}^2 \Phi = \Delta \Phi = 0 \,.
\label{eq:solar_laplace}
\end{align}
The PFSS model uses one parameter, the source-surface radius $R_{\mathrm{ss}}$, which delimits the domain in which the magnetic field dominates the plasma.
Beyond this source surface, the plasma becomes super-alfvénic and the magnetic field is passively advected outwards in it. 
The source-surface radius is set to $R_{\mathrm{ss}} = 2.5\,R_{\odot}$, which is a commonly used value and has also been tested in \cite{Tibet2013}.
\subsubsection{CSSS model}
The current-sheet source-surface (CSSS) model \citep{Zhao1995} is based on the magnetohydrostatic equation
\begin{align}
\vec{0} = \vec{j} \times \vec{B} - \vec{\nabla} p + \rho \vec{g} \,,
\label{eq:mhs}
\end{align}
which balances the Lorentz force, the gradient of the plasma pressure $p$, and the gravitational acceleration $\vec{g}$ that acts on the plasma density $\rho$.

The CSSS model is based on the solution presented in \cite{Bogdan1986} and uses three parameters: the source-surface radius $R_{\mathrm{ss}}$, the cusp radius $R_{\mathrm{cp}}$, and the length scale $l_{\mathrm{a}}$ of horizontal currents. 
While $R_{\mathrm{ss}}$ is set to $2.5\,R_{\odot}$ and has the same meaning as in the PFSS model, $R_{\mathrm{cp}}$ is the radius where magnetic field lines become closed.
$R_{\mathrm{cp}}$ is set to $1.7 \, R_{\mathrm{\odot}}$, which is a typical height for coronal streamers. 
Above $R_{\mathrm{cp}}$, magnetic field lines are assumed to be open. 
The length scale $l_{\mathrm{a}}$ of horizontal currents is set to $1\,R_{\odot}$.
\subsubsection{Parker spiral}
The heliospheric magnetic field is implemented using the model first developed by Parker \citep{Parker1958} (cf. \cite{Owens2013} for a review). The Parker spiral in general has a footpoint as an inner boundary of the description of the field, which for the Sun is assumed to be the photosphere \citep{Owens2013}.
While the magnetic field at the footpoints of the spiral is determined by using the coronal models described above, the radial velocity $V_r$ of the solar wind, which in the frozen-magnetic-flux model determines the radial component of the magnetic field, is modeled as a semi-linear approximation of the Parker \citep{Parker1958} isothermal solar wind profile:
\begin{align}
V_r (r) = \begin{cases}
C \cdot \left(r/R_{\odot}\right) & r < R_{\rm c} \\
V_0 & r \geq R_{\rm c} \,,
\end{cases}
\label{eq:sim_solar_wind_speed}
\end{align}
with the slope $C = V_0 \cdot(R_{\odot}/R_{\rm c}) = \SI{20}{\kilo\meter\per\second}$ and the critical radius $R_c = 22.5\,R_{\odot}$.
Beyond $R_{\rm c}$, the radial velocity is assumed constant with a value of $V_0 = \SI{450}{\kilo\meter\per\second}$, which is a typical value for the radial solar wind velocity at \SI{1}{\astronomicalunit} (cf. \cite{Owens2013}).
\subsection{Coordinate transformation for signal simulation\label{coordinate_transform_sim:sec}}
For a proper description of the propagation, coordinates need to be transformed into ecliptic coordinates before starting the propagation around the Sun, which adds an additional transformation step.
The relative coordinates with respect to the Sun's position, $\Delta \lambda$ and $\Delta \beta$ are given by
\begin{eqnarray}
\Delta\lambda&=&\lambda_{\rm cr}-\lambda_{\odot}\\
\Delta \beta &=& \beta_{\rm cr}-\beta_{\odot}\,,
\end{eqnarray}
with $(\lambda_{\odot},\,\beta_{\odot})$ as the position of the Sun in ecliptic coordinates and $(\lambda_{\rm cr},\,\beta_{\rm cr})$ as the position of the detected cosmic ray in ecliptic coordinates.
These relative coordinates can then be transformed into  quasi-Cartesian coordinates as follows:
\begin{eqnarray}
x'&=&\cos\beta_{\rm cr}\cdot \Delta \lambda\\
y'&=&\Delta \beta\,.
\end{eqnarray}
Here, $x'$ is corrected for the spherical distortion by the $\cos\beta_{\rm cr}$ factor, just as it is done for the Moon in equatorial coordinates. While Earth’s axial tilt
of about $23.4^{\circ}$ is taken into account by transforming from equatorial to ecliptic coordinates,
the tilt of the Sun’s rotational (and magnetic) axis with respect to the ecliptic of about $7.25^{\circ}$
 is neglected in this approach, as it is significantly smaller than the Earth's axial tilt with respect to the ecliptic. 
 
 Finally, the coordinates are transformed back into  the equatorial system that is used for the data analysis of Moon and Sun, and also for the simulation of the Moon data.
\subsection{\label{ssec:analysis_energy}Energy reconstruction}
For studying the cosmic-ray Sun shadow at different energies, the data are divided into three energy bands.
This is achieved by using an energy-correlated observable, $q_{\mathrm{dir}}$, which represents the charge deposited in direct hits. These direct hits are defined as not having undergone significant scattering in the ice from the point of their emission, thus providing more accurate timing information. For the observable $q_{\mathrm{dir}}$, the sum of charge deposited in all DOMs that are hit within a time window of (\SI{-15}{\nano\second}, \SI{+75}{\nano\second}) around the first geometrically possible arrival of a Cherenkov photon in a DOM is given in units of photo-electrons (p.e.).

The three bins are defined as $q_{\mathrm{dir}} < 18\,\mathrm{p.e.}$, $18 \leq q_{\mathrm{dir}} \leq 30\,\mathrm{p.e.}$, and $q_{\mathrm{dir}} > 30\,\mathrm{p.e.}$, resulting in an approximately equal number of events in each energy bin, see Table \ref{tab:energy} for details. These three sub-samples have median primary energies of \SI{40}{\TeV}, \SI{55}{\TeV}, and \SI{100}{\TeV} as shown in Figure \ref{fig:EnergyDist}.

\begin{table}[htbp]
\centering
\caption[Energy-dependent analysis: three energy bins]{Summary of the parameters for the three energy bins of the analysis}
\label{tab:energy} 
\begin{tabular}{@{}lrr@{}}
\toprule
$q_{\rm dir}$/p.e.&range (68\%)/TeV &$E_{\rm median}/$TeV \\
\midrule
 $<18$	& $(12-160)$&$40$ \\
$18-30$ &$(15-260)$ &$55$ \\
$>30$   &$(21-630)$ &$100$\\
\bottomrule
\end{tabular}
\end{table}
\section{\label{sec:results}Results}
\subsection{Shadow maps}
\begin{figure*}[htbp]
\centering
\includegraphics[trim=0 0 2.6cm 0, clip,    				height=4.5cm]{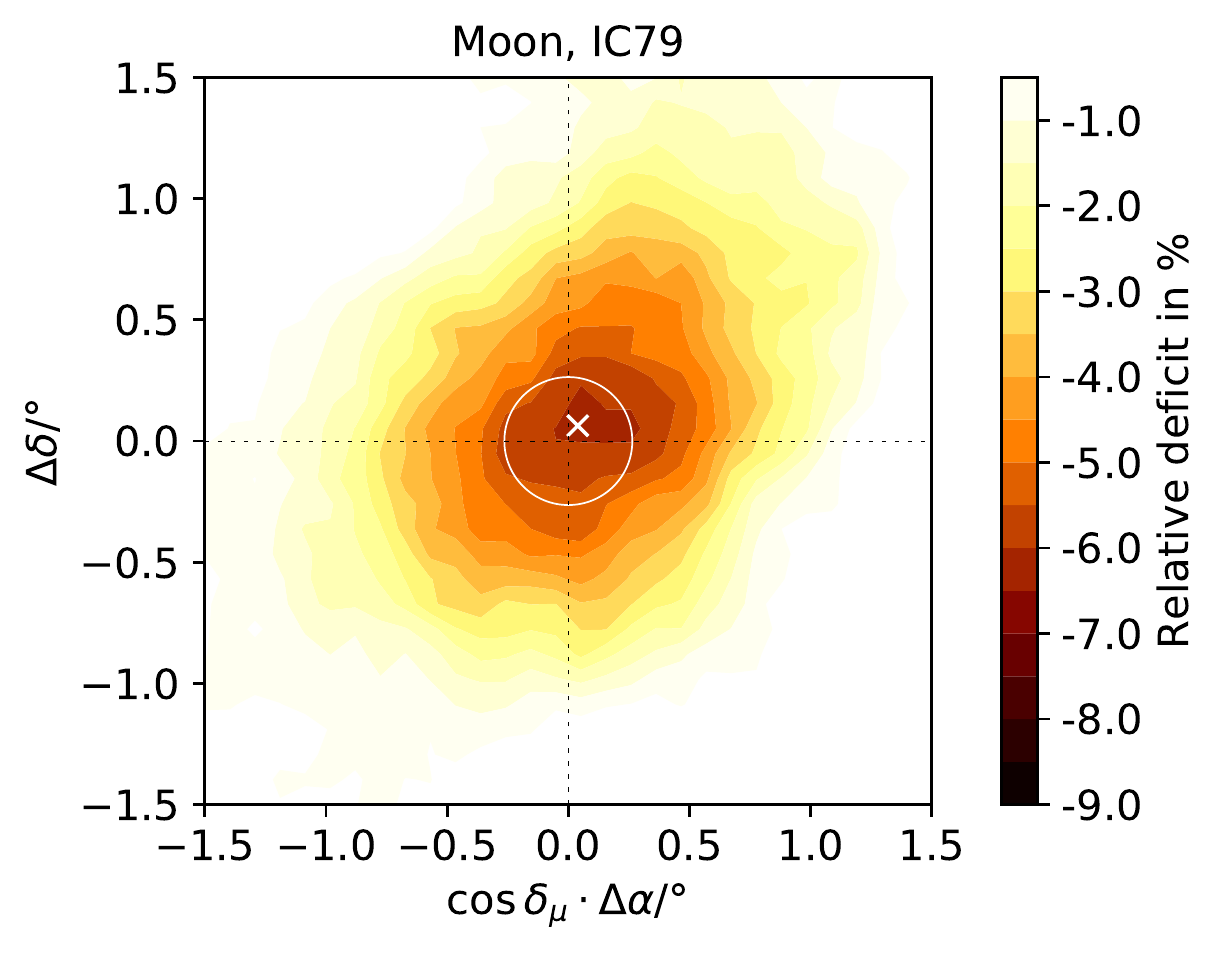}
\includegraphics[trim=1.9cm 0 2.6cm 0, clip, 			height=4.5cm]{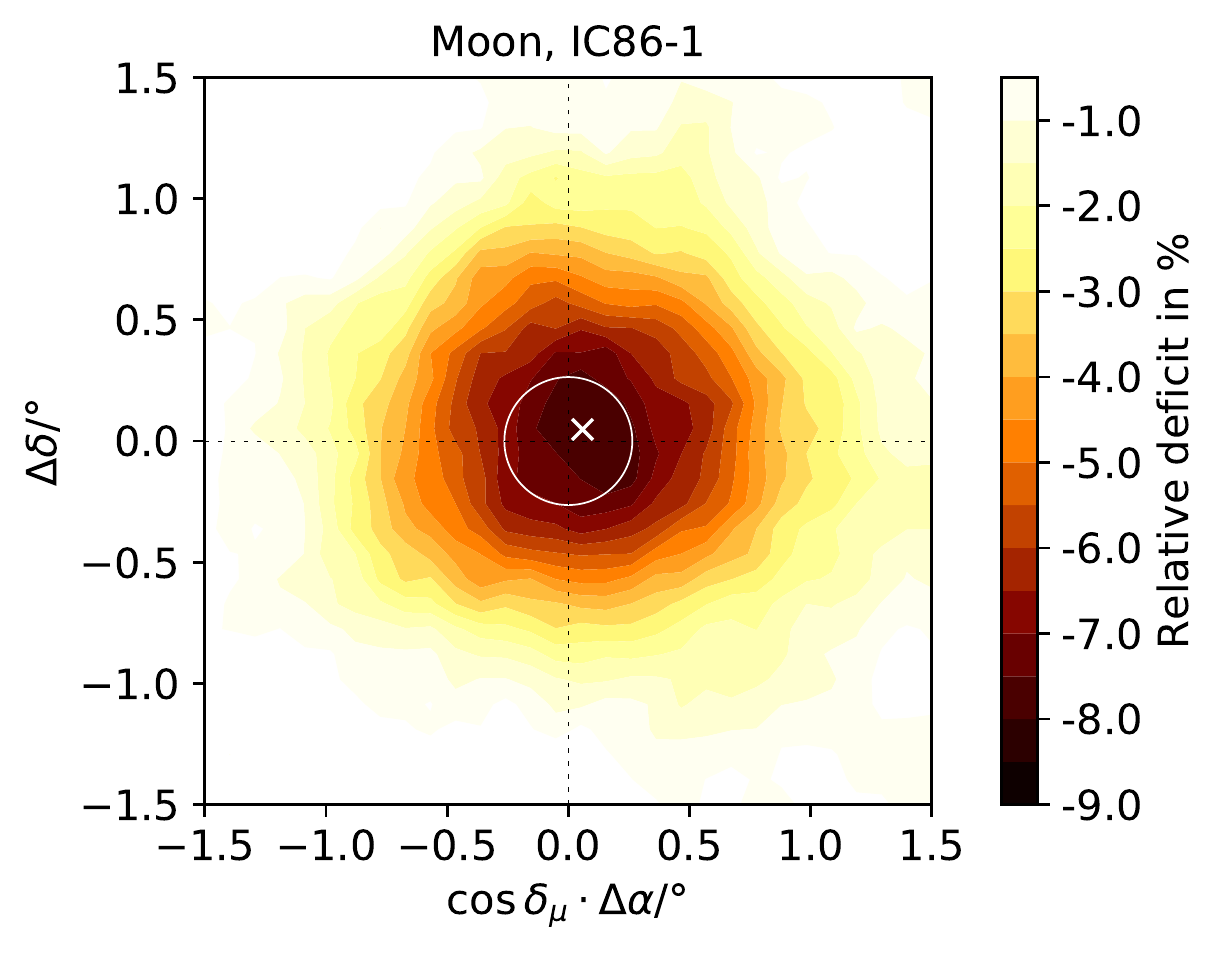}
\includegraphics[trim=1.9cm 0 2.6cm 0, clip, 					height=4.5cm]{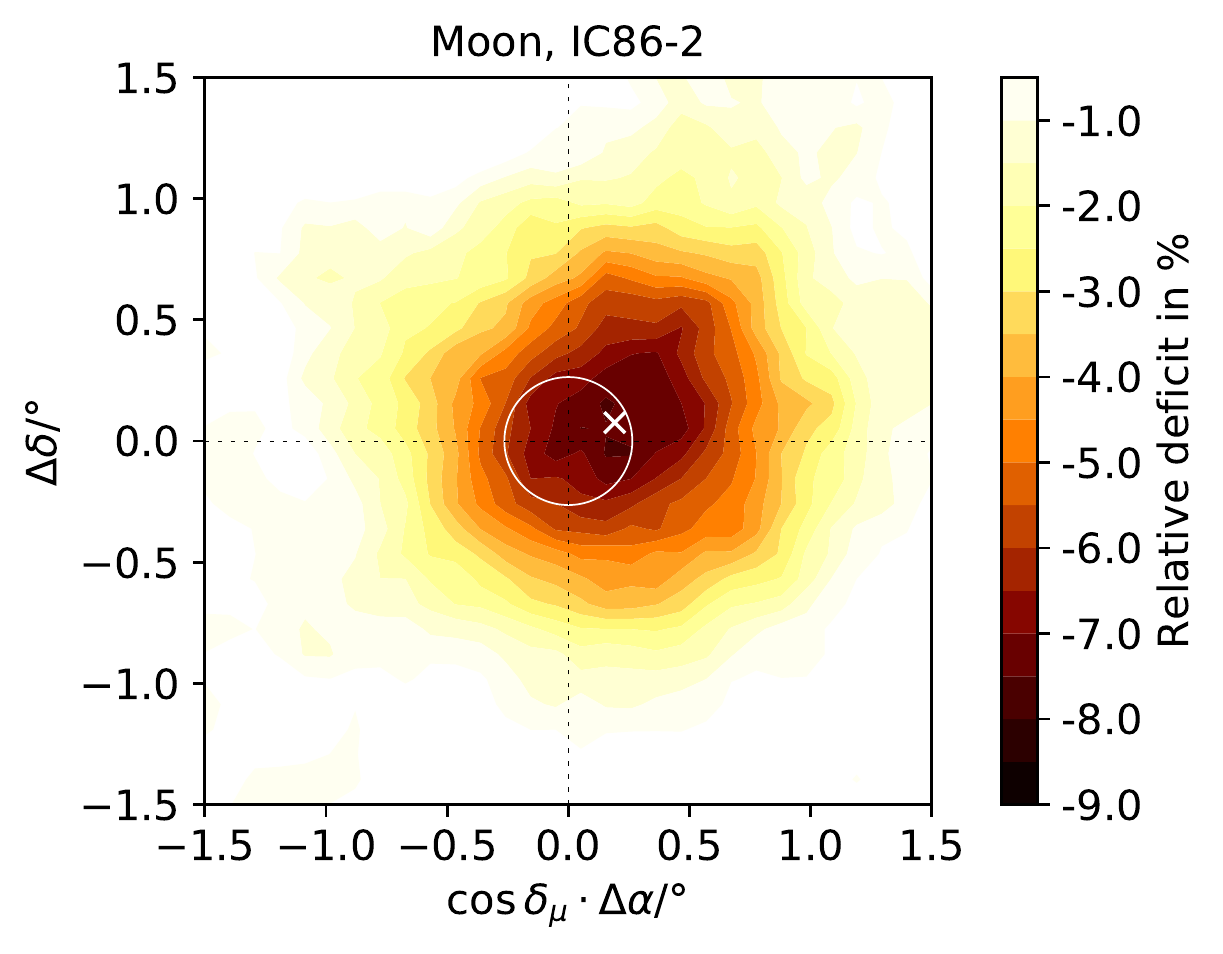}
\includegraphics[trim=1.9cm 0 0 0, clip,   					height=4.5cm]{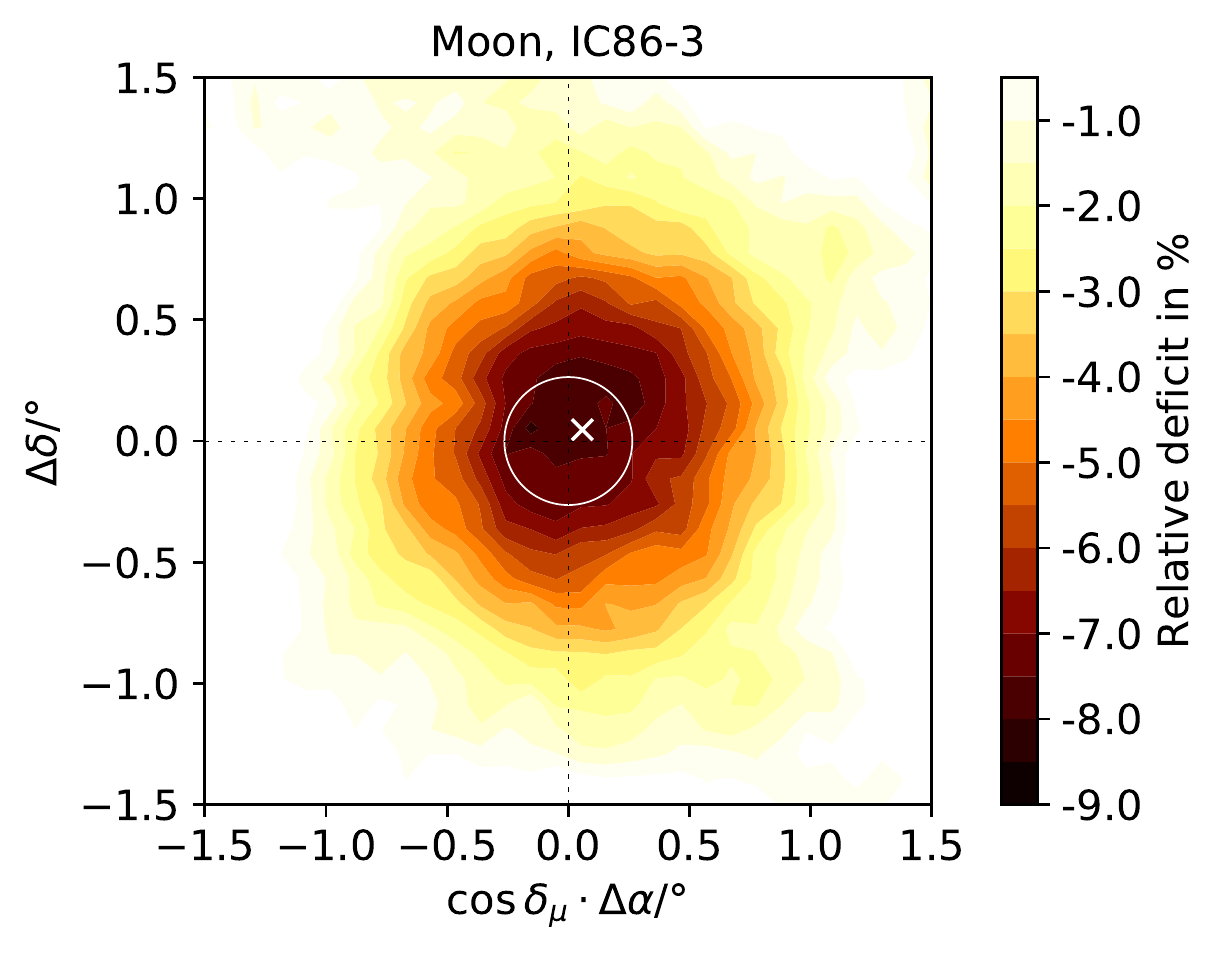}

\includegraphics[trim=1.9cm 0 2.6cm 0, clip,    			height=4.5cm]{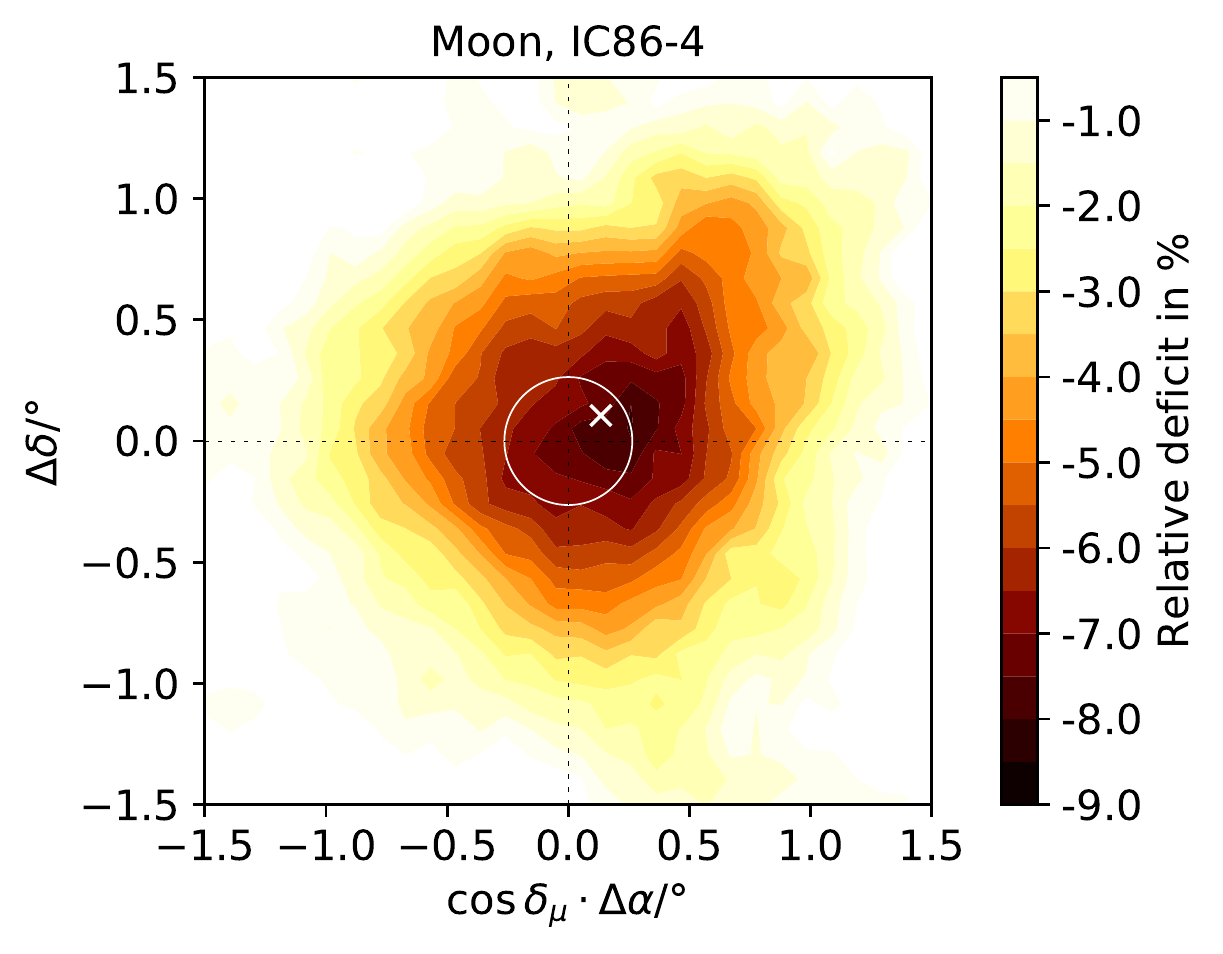}
\includegraphics[trim=1.9cm 0 2.6cm 0, clip, 					height=4.5cm]{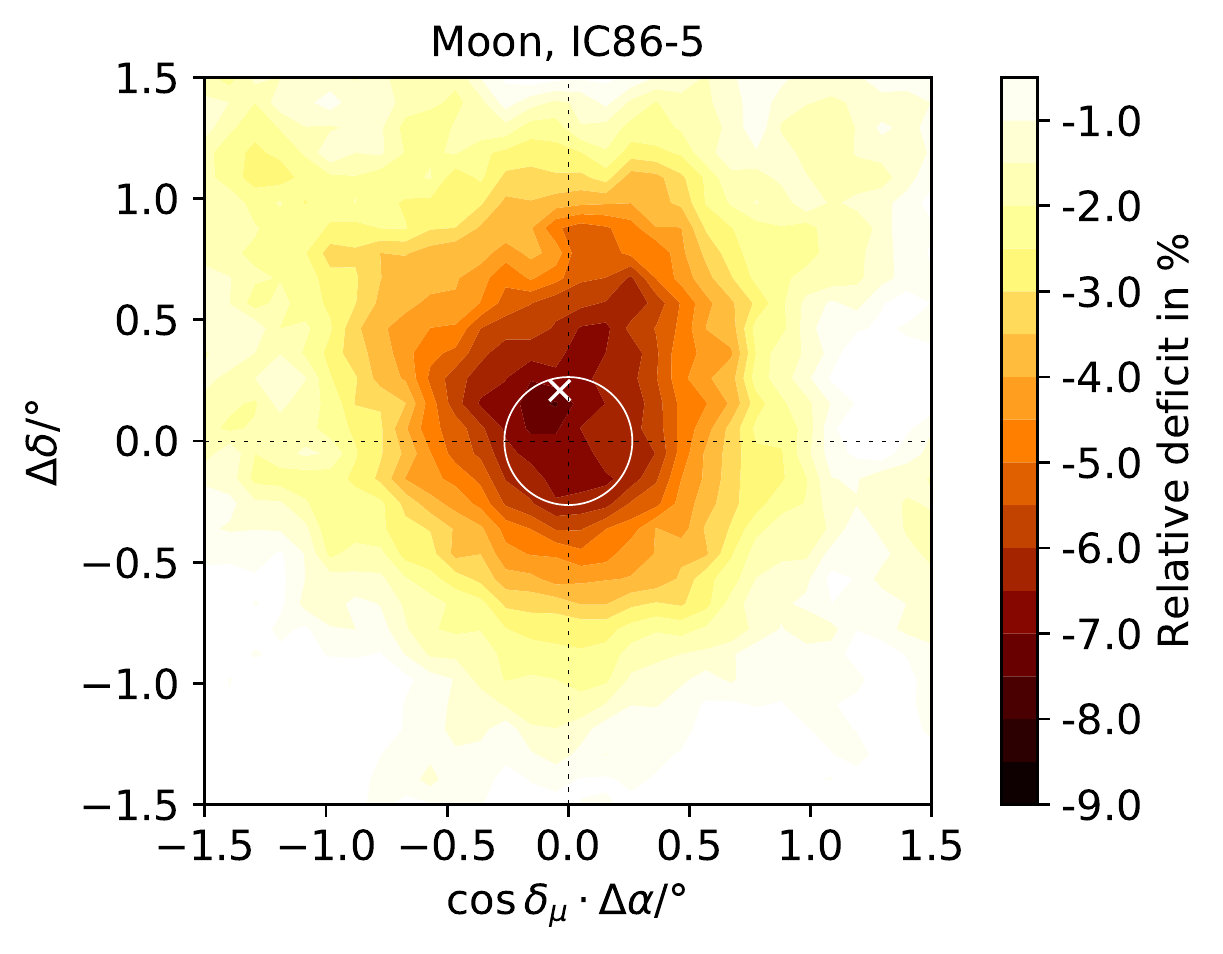}
\includegraphics[trim=1.9cm 0 0 0, clip,        					height=4.5cm]{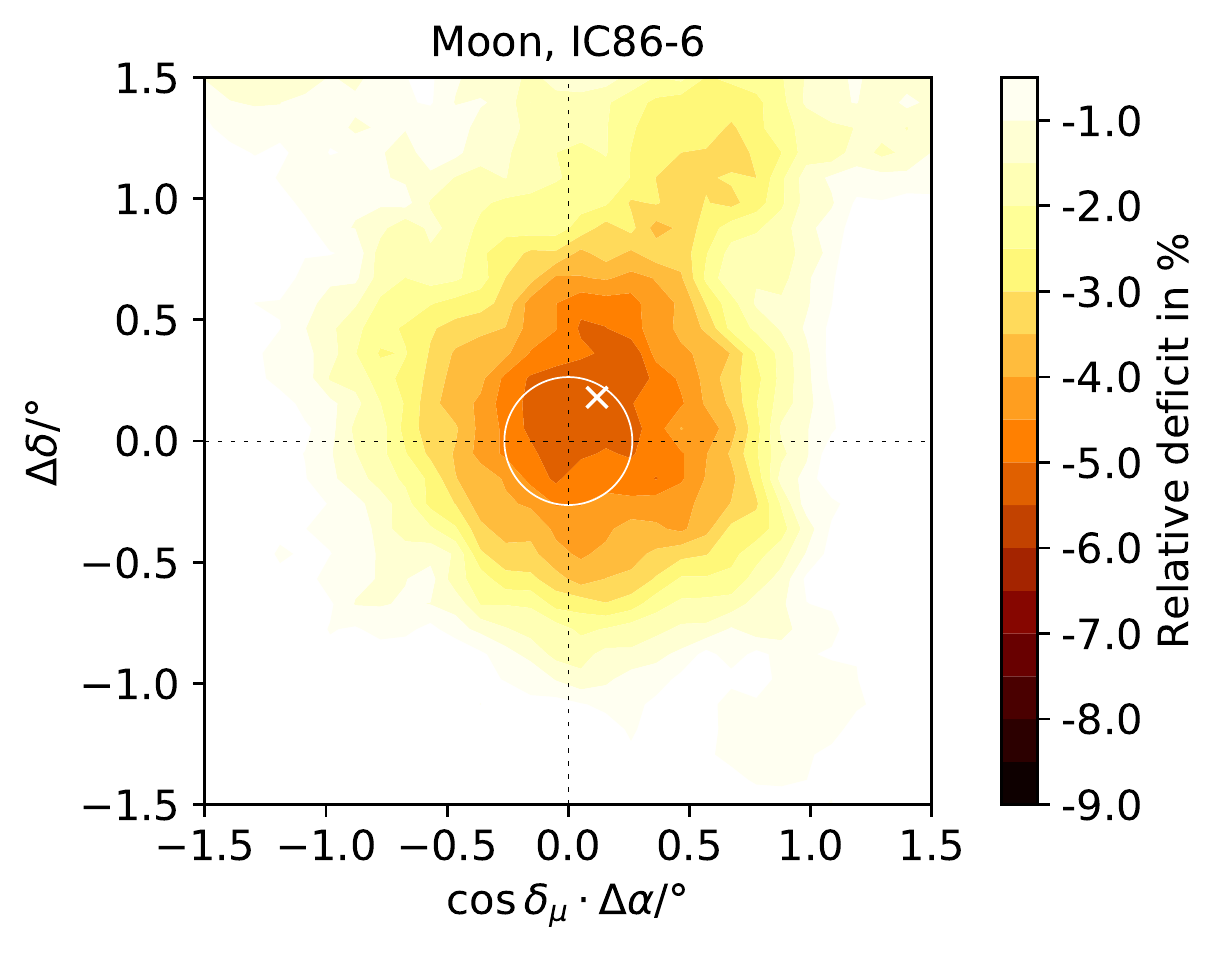}
\caption[Moon shadow from IC79 to IC86-6: 2D, box-car-smoothed]{Boxcar-smoothed two-dimensional contour map of the Moon shadow for the years IC79 to IC86-6 showing the computed center of gravity of the shadow as a white cross. The white circle indicates the seven-year mean of the weighted average of the angular Moon radius.}
\label{fig:shadows_moon_2d}
\end{figure*}

The shadow maps for the Moon and the Sun as a result of this analysis are presented here for each IceCube season. Figure \ref{fig:shadows_moon_2d} shows the cosmic-ray shadow of the Moon. Each panel shows one year of data, starting with the earliest season 2010/2011 (IC79) on the top-left, followed by data for the seasons 2011/2012 (IC86-1), 2012/2013 (IC86-2) and 2013/2014 (IC86-3) in the top row and 2014/2015 (IC86-4), 2015/2016 (IC86-5) and 2016/2017 (IC86-6) in the bottom row from the left to the right. 
Data have been smoothed with the boxcar average algorithm, where the smoothed relative deficit in each bin $(i,\, j)$ is determined as the average of all bins with centers within a certain angular
distance around the center of bin $(i,\,j)$. Here, this smoothing radius is set to $0.7^{\circ}$,
which approximately corresponds to the median angular resolution of the simulation
sample and yields a reasonable balance between angular resolution and statistical uncertainty. 

\begin{figure*}[htbp]
\centering
\includegraphics[trim=0 0 2.6cm 0, clip,    				height=4.5cm]{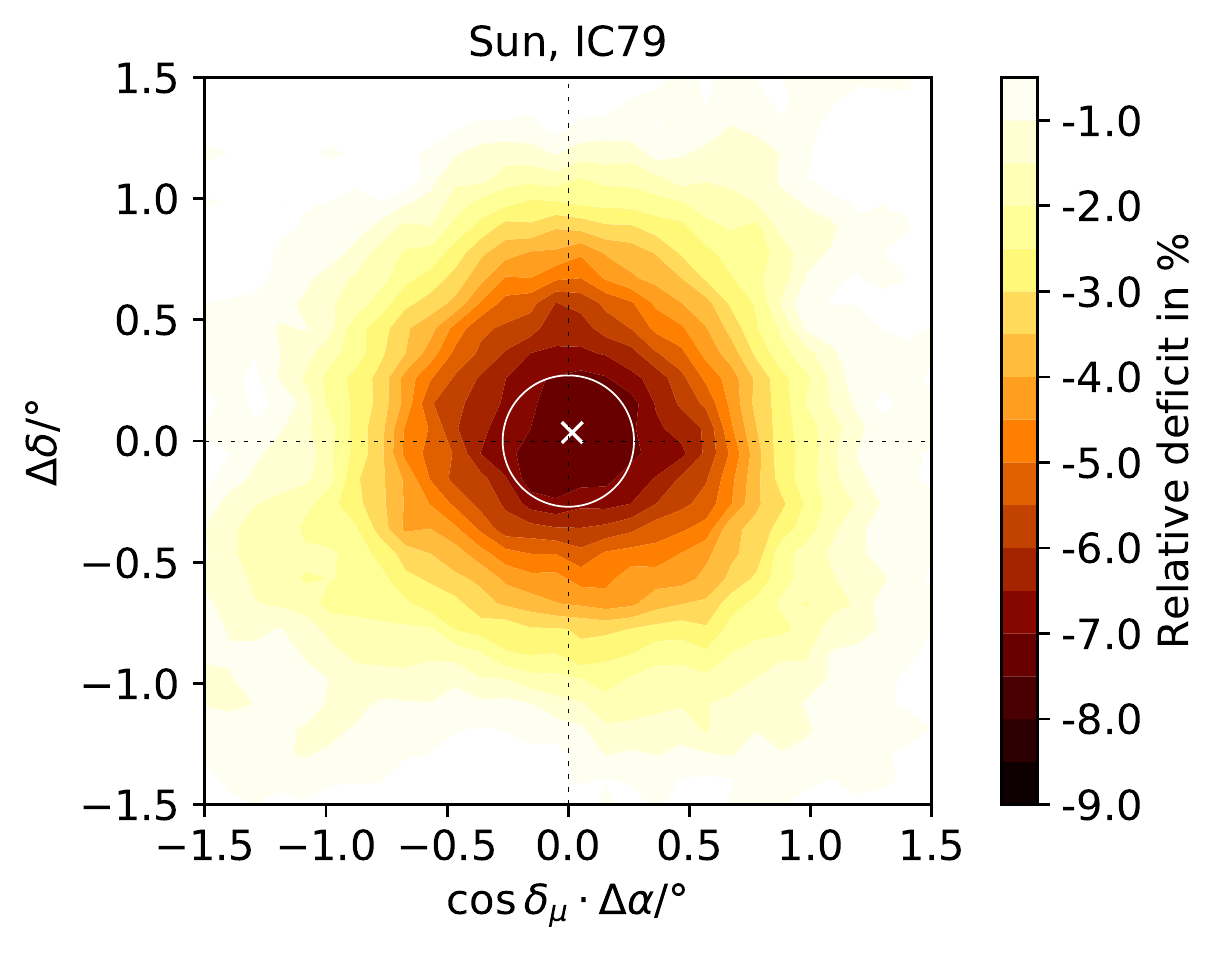}
\includegraphics[trim=1.9cm 0 2.6cm 0, clip, 			height=4.5cm]{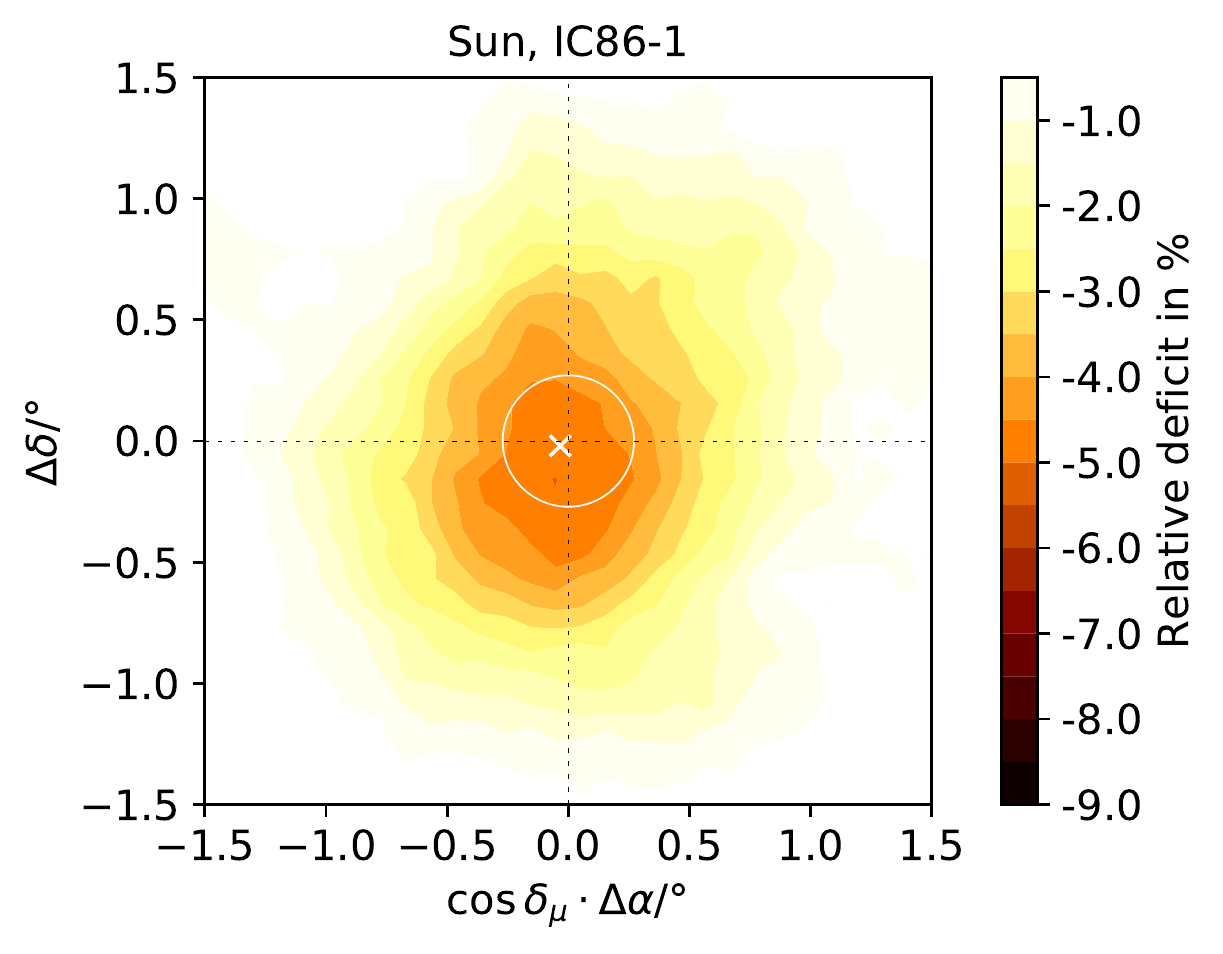}
\includegraphics[trim=1.9cm 0 2.6cm 0, clip, 					height=4.5cm]{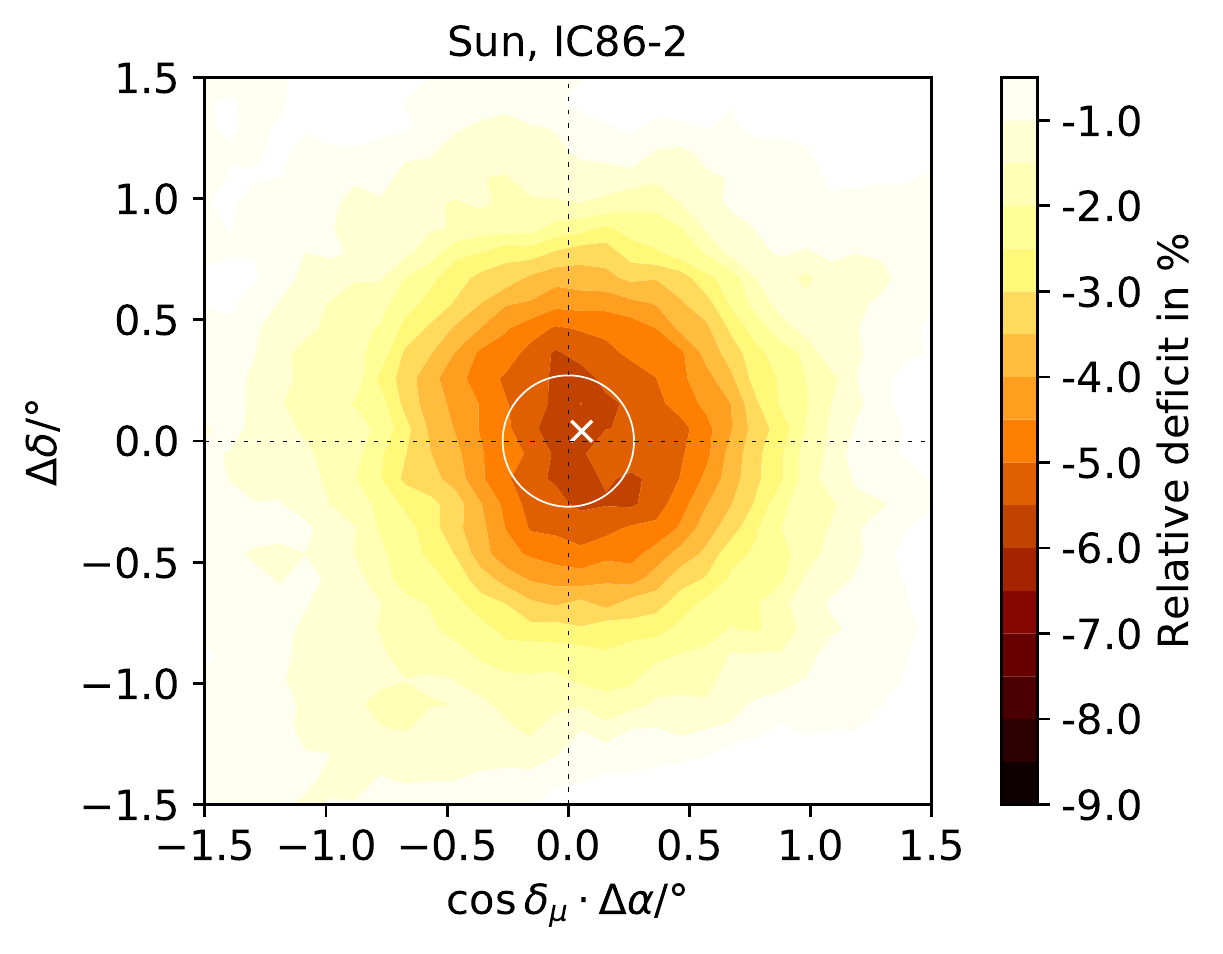}
\includegraphics[trim=1.9cm 0 0 0, clip,   					height=4.5cm]{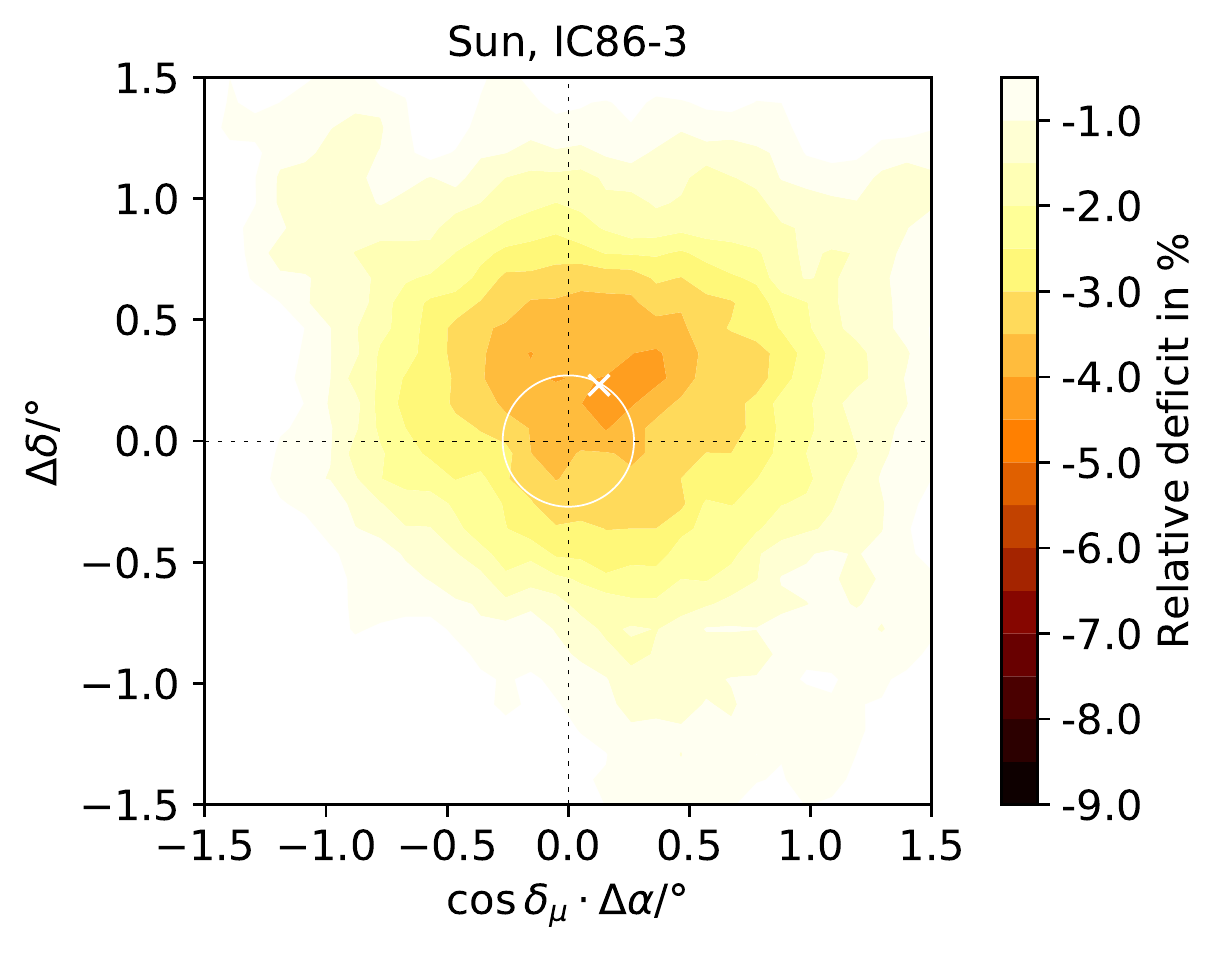}

\includegraphics[trim=1.9cm 0 2.6cm 0, clip,    			height=4.5cm]{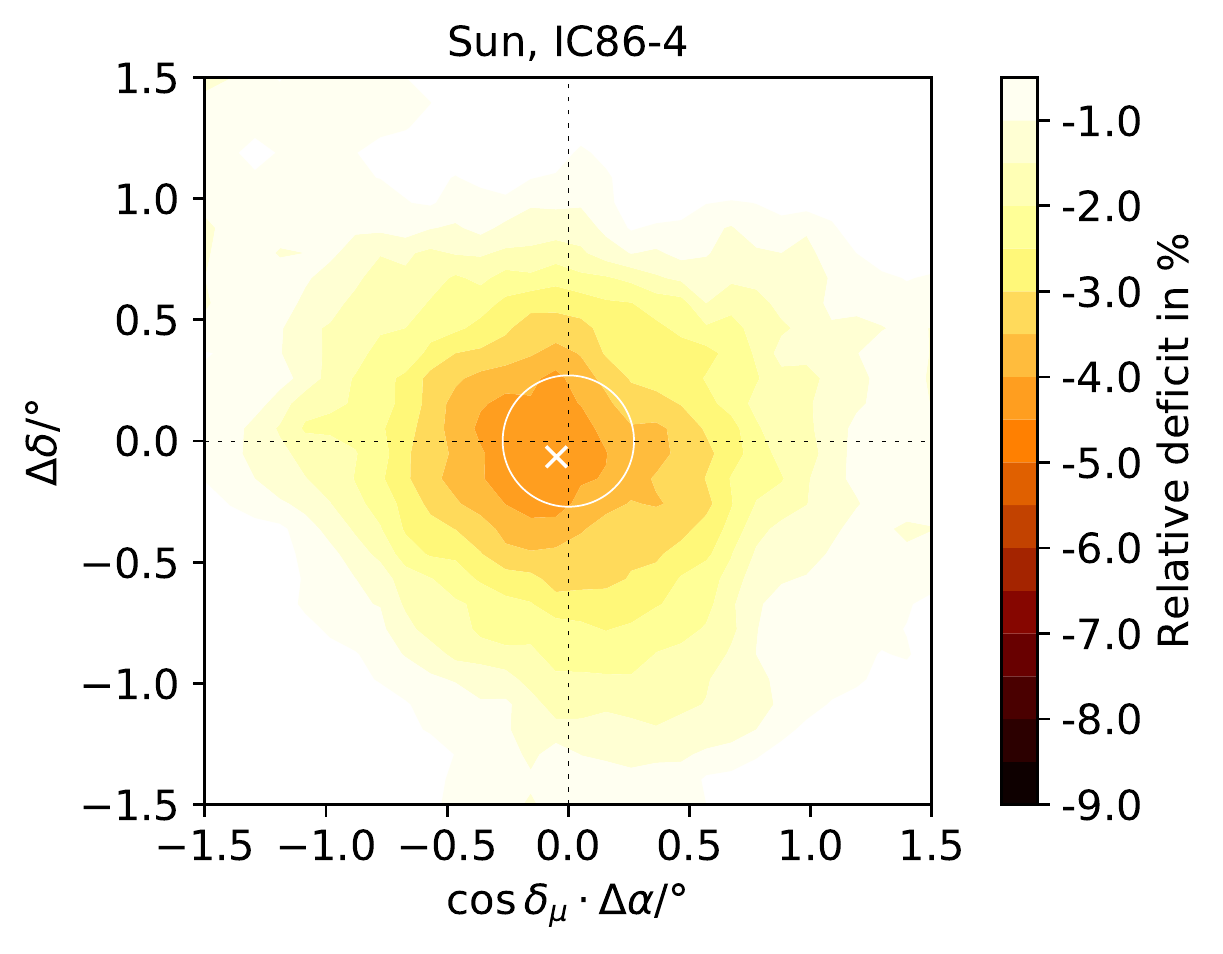}
\includegraphics[trim=1.9cm 0 2.6cm 0, clip, 					height=4.5cm]{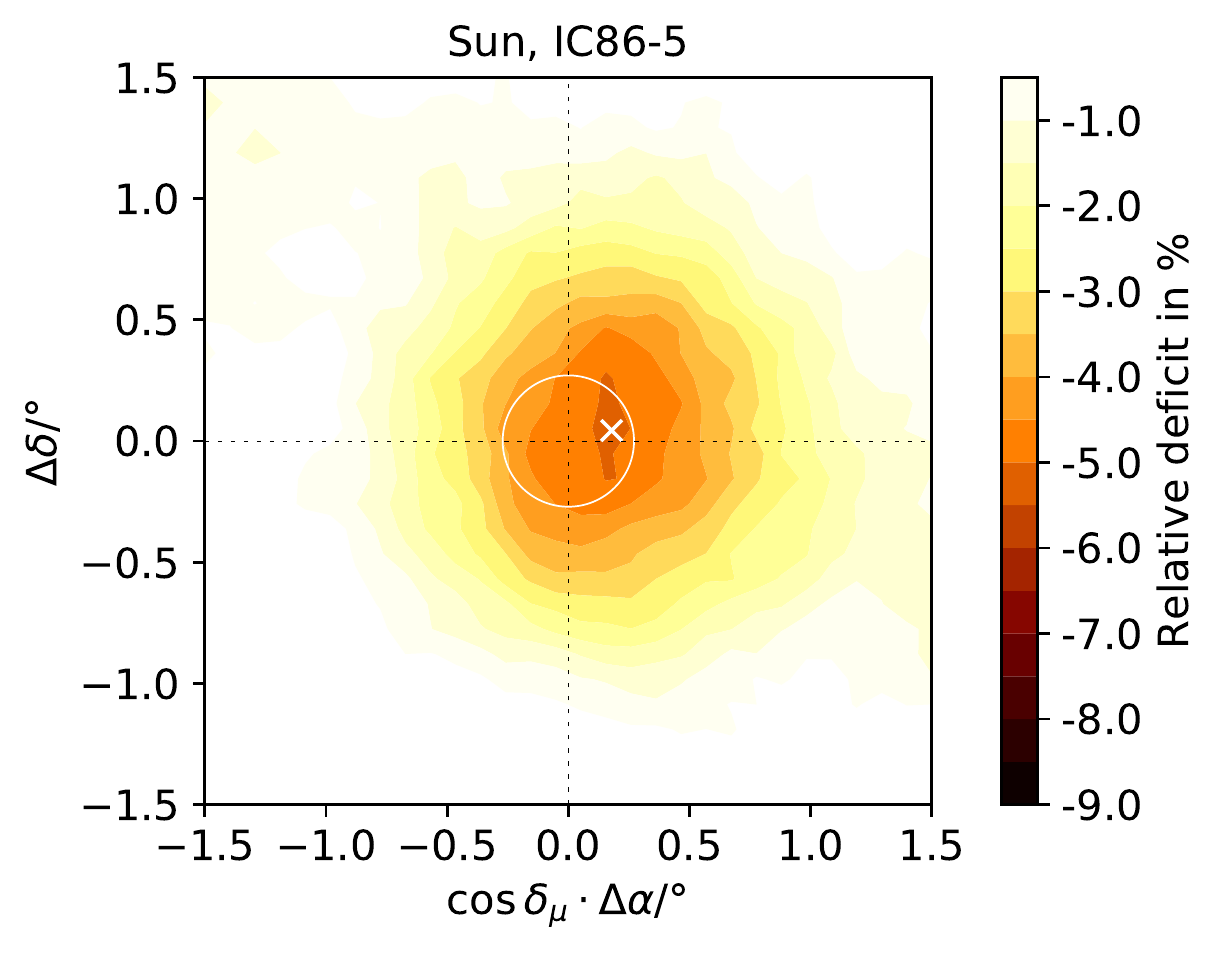}
\includegraphics[trim=1.9cm 0 0 0, clip,        					height=4.5cm]{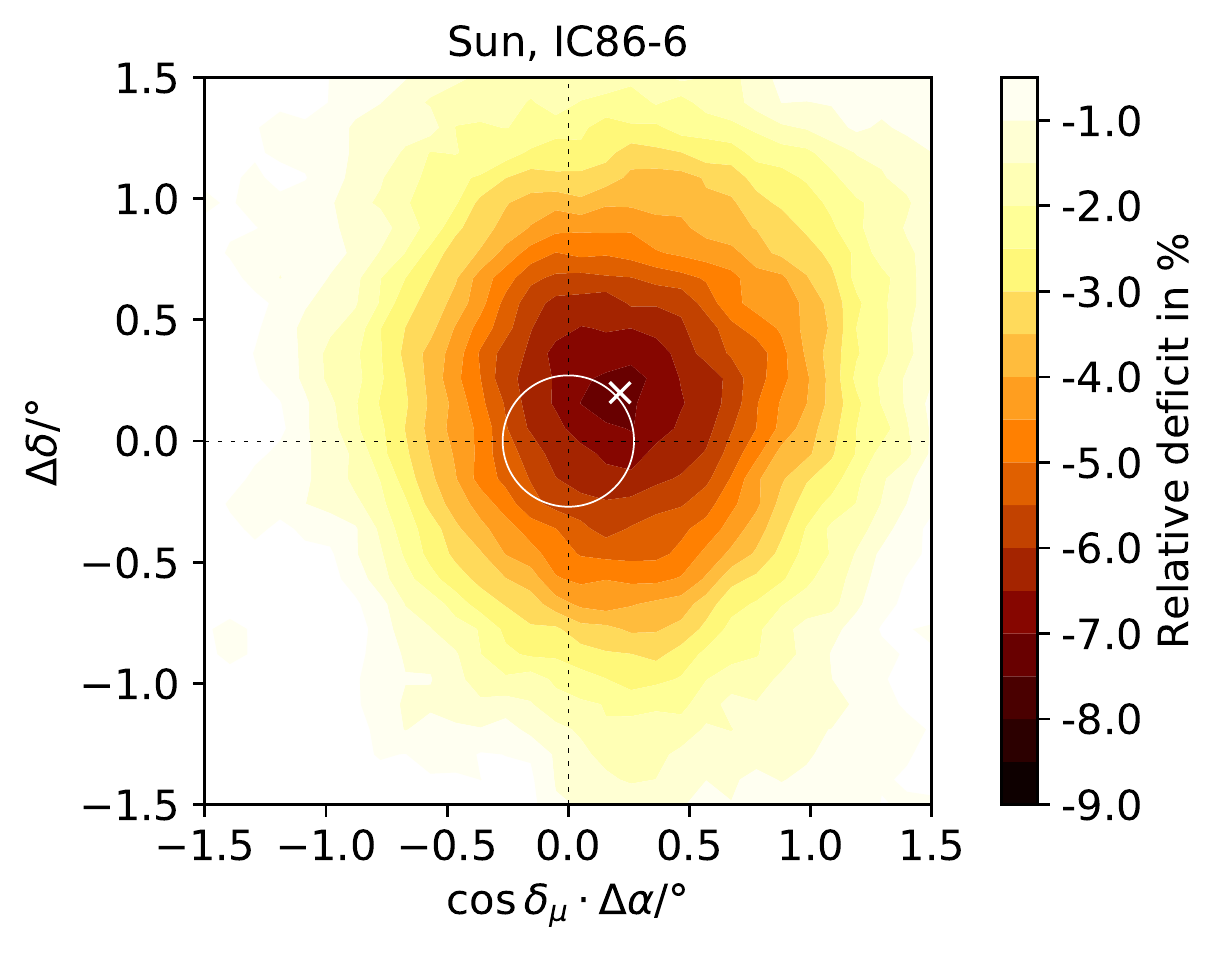}
\caption[Sun shadow from IC79 to IC86-6: 2D, box-car-smoothed]{Boxcar-smoothed two-dimensional contour map of the Sun shadow for the years IC79 to IC86-6 showing the computed center of gravity of the shadow as a white cross. The white circle indicates the weighted average of the angular Sun radius.}
\label{fig:shadows_sun_2d}
\end{figure*}

The figure shows that the relative deficit reaches a depth larger than $8.0\%$. Figure \ref{fig:shadows_sun_2d} shows the corresponding pictures for the location of the Sun. 

The significance of the shadowing effect (cf. Section \ref{sssec:analysis_method_num}) is found to fall between  $7.5\sigma$ and $14.2\sigma$ for the Moon and between approximately $9.5\sigma$ and $16.9\sigma$ for the Sun (see Table \ref{tab:sigma}). The reason for the higher significance of the Sun shadow is its larger data sample.
An interpretation of these figures, and in particular a quantification of a possible temporal change in the shadow, will be given in the next section.

\begin{table*}[htbp]
\centering
\caption[Relative Deficit \& Significance of the shadow deficit]{Relative deficit (RD) and Li-Ma significance $S$ (cf. Section \ref{sssec:analysis_method_num}) for Moon and Sun shadows.}
\label{tab:sigma} 
\begin{tabular}{@{}lcccccccc@{}}
\toprule
&&IC79&IC86-1&IC86-2&IC86-3&IC86-4&IC86-5&IC86-6  \\
\midrule
\multirow{2}{*}{Moon} 
& RD in \percent&$4.0\pm0.4$&$5.0\pm0.4$&$4.5\pm0.4$&$5.6\pm0.5$&$5.3\pm0.6$&$4.9\pm0.6$&$3.7\pm0.5$\\
& $S$ in $\sigma$ 	&$11.2$&$14.2$&$12.1$&$10.0$&$9.4$&$9.5$ &$7.5$\\
\midrule
\multirow{2}{*}{Sun} 
& RD in \percent&$5.1\pm0.4$&$3.3\pm0.3$&$4.1\pm0.3$&$3.1\pm0.3$&$2.8\pm0.3$&$3.5\pm0.3$&$5.2\pm0.3$\\
& $S$ in $\sigma$&$14.0$&$11.4$&$13.0$&$9.5$&$10.1$&$12.1$ &$16.9$ \\ 
\bottomrule
\end{tabular}
\end{table*}
\clearpage
\subsection{\label{ssec:results_disk}Comparison to lunar/solar disk}
\begin{figure}[htbp]
    \centering
    \includegraphics[width=\linewidth]{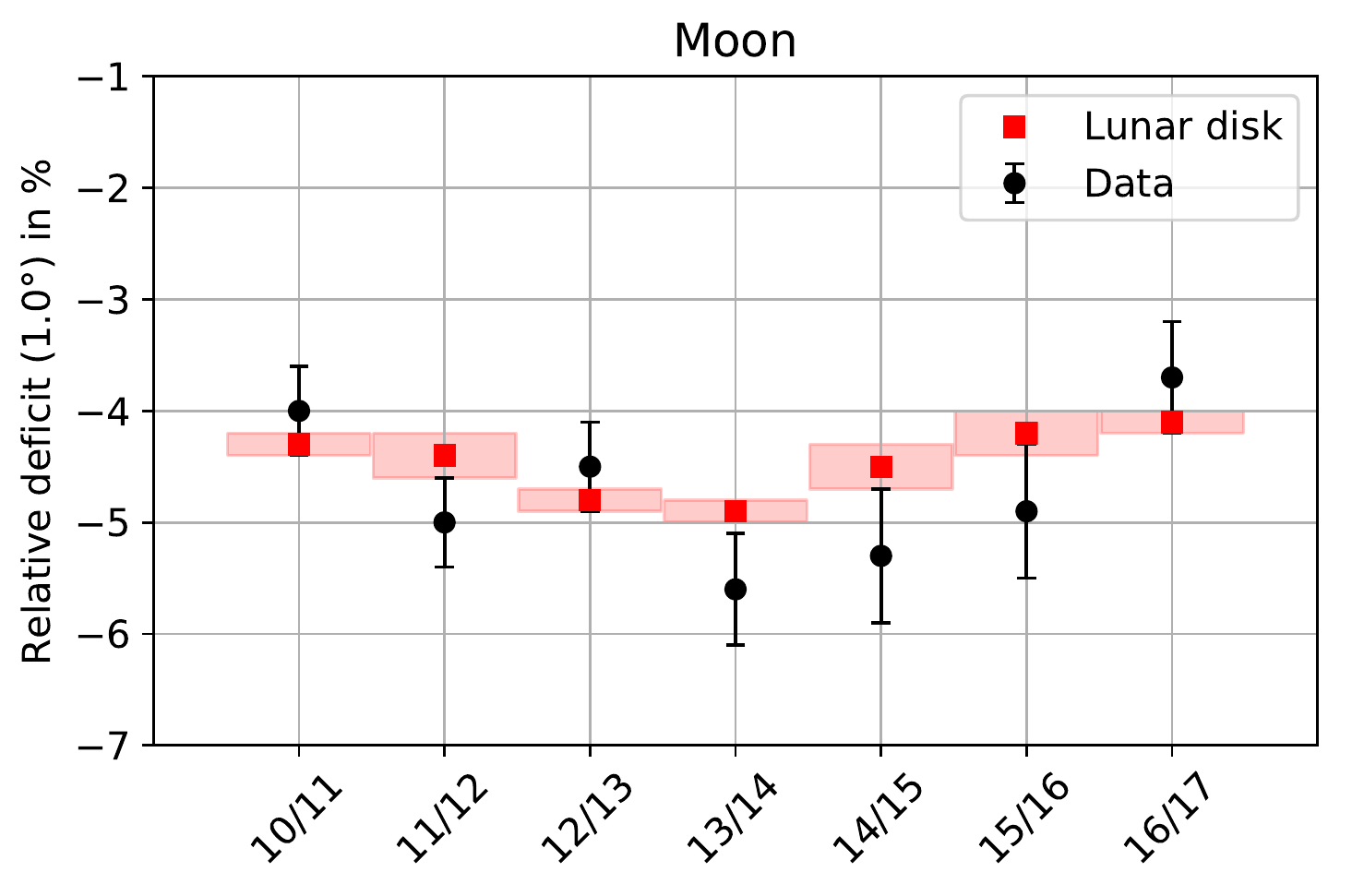}
    \caption{Comparison of measured relative deficit due to the Moon shadow and relative deficit expected from shadowing by the lunar disk.}
    \label{fig:MoonDisk}
\end{figure}

\begin{figure}[htbp]
    \centering
    \includegraphics[width=\linewidth]{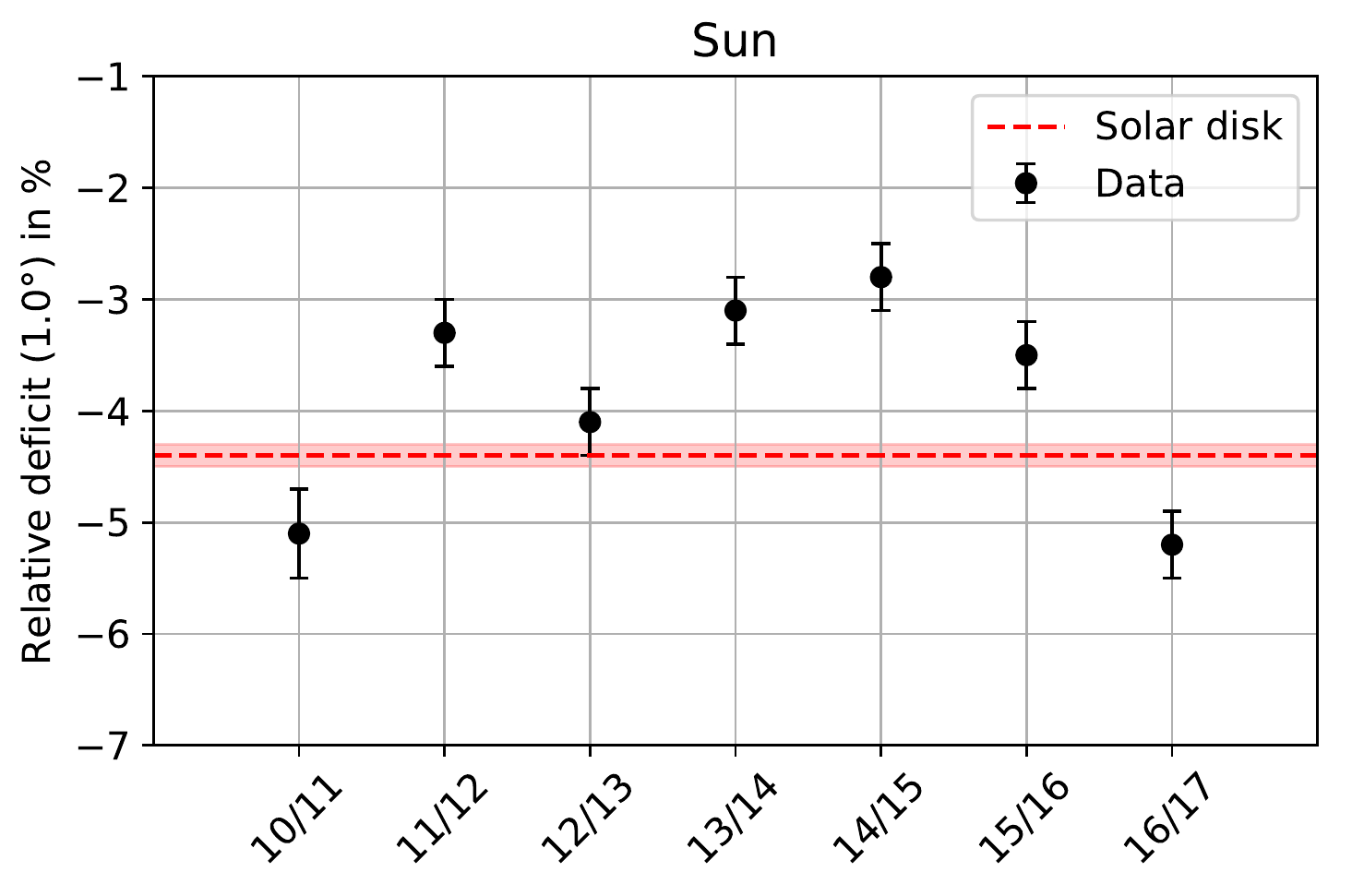}
    \caption{Comparison of measured relative deficit due to the Sun shadow and relative deficit expected from shadowing by the solar disk.}
    \label{fig:SunDisk}
\end{figure}

As described in Section \ref{sssec:analysis_method_num}, the relative deficit within a \ang{1.0}-circle around the center of gravity of the shadow is used for quantifying the deficit of cosmic-ray induced muon events due to the Moon and Sun shadows. Figures \ref{fig:MoonDisk}, \ref{fig:SunDisk}, \ref{fig:RelDefSA}, \ref{fig:RelDefSACorr}, \ref{fig:SunModels}, and \ref{fig:EnergyDep} thus use this quantity.
In Figure \ref{fig:MoonDisk} the observed relative deficit due to the cosmic-ray Moon shadow is compared to the relative deficit expected due to geometrical shadowing of the Moon.
The simulations show the same slight dip as the data. 
The reason for this dip is the slightly different distance between Moon and Earth, which changes the angular radius and hence the shadowed solid angle. 
Additionally, the average declination of the event sample is slightly different for each year. 
Both effects are accounted for in the simulations shown in Figure \ref{fig:MoonDisk}.
{}

In Figure \ref{fig:SunDisk} the observed relative deficit due to the cosmic-ray Sun shadow is compared to the relative deficit expected due to geometrical shadowing of the Sun.
There is no substantial variation in the distance between Sun and Earth for the observation period November through February.
Also, the average declination of the data sample is essentially the same each year. 
Thus, the expected relative deficit due to the geometrical shadowing by the solar disk is the same every year and amounts to \SI{4.4+-0.1}{\percent}. 
{}

In Table \ref{tab:sim_disk_comparison}, the reduced $\chi^2$, $p$-value, and significance $S$ of a $\chi^2$-test of the observed Moon and Sun shadows and the expectation from the lunar/solar disk are given. 
With a $p$-value of \SI{32}{\percent} the Moon shadow shows reasonable agreement with the expectation from the lunar disk.
The Sun shadow, on the other hand, is incompatible with the expectation from the solar disk with a statistical significance of about $7$ standard deviations. 
\begin{table}[htbp]
\centering
\caption[Results: comparison with lunar/solar disk]{Reduced $\chi^2$, $p$-value and, significance of the comparison between the measured Moon and Sun shadows with the expectation from the lunar and solar disk.}
\label{tab:sim_disk_comparison} 
\begin{tabular}{@{}lccc@{}}
\toprule
& $\chi^2 / n_{\mathrm{dof}}$ & $p$ & $S$ in $\sigma$ \\
\midrule
Moon 	& $8.2/7 \approx 1.2$ 		& \num{0.32} 			& \num{1.0} \\
Sun 		& $72.9/7 \approx 10.4$ 	& \num{3.9e-13} 	& \num{7.3} \\
\bottomrule
\end{tabular}
\end{table}
%
\subsection{\label{ssec:results_cycle}Comparison to solar cycle}
\begin{figure}[htbp]
    \centering
    \includegraphics[width=\linewidth]{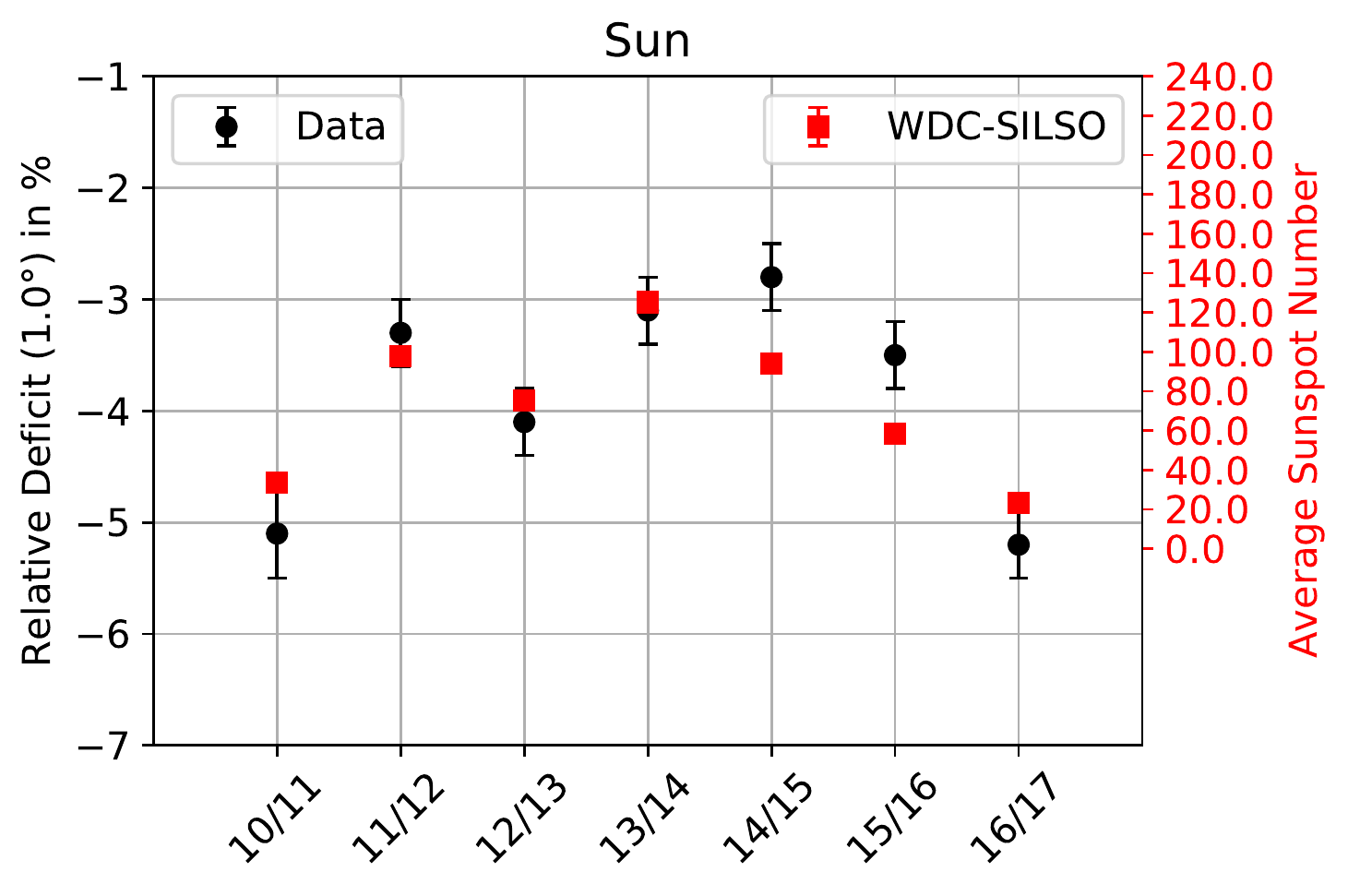}
    \caption{Comparison of measured relative deficit due to the Sun shadow and average sunspot number as a tracer for solar activity.}
    \label{fig:RelDefSA}
\end{figure}

As a first observational test of a connection between magnetic solar activity and the Sun shadow, the temporal variation of the cosmic-ray Sun shadow is compared to the average \textit{International Sunspot Number} obtained from \cite{SIDC2019}. 
A similar comparison has already been performed in \cite{MoonSunApj2019} for five years of data, and a correlation has been found to be likely. 
In Figure \ref{fig:RelDefSA}, the relative deficit due to the Sun shadow is shown together with the sunspot number (averaged over the relevant months) between November 2010 and February 2017.

\begin{figure}[htbp]
    \centering
    \includegraphics[width=\linewidth]{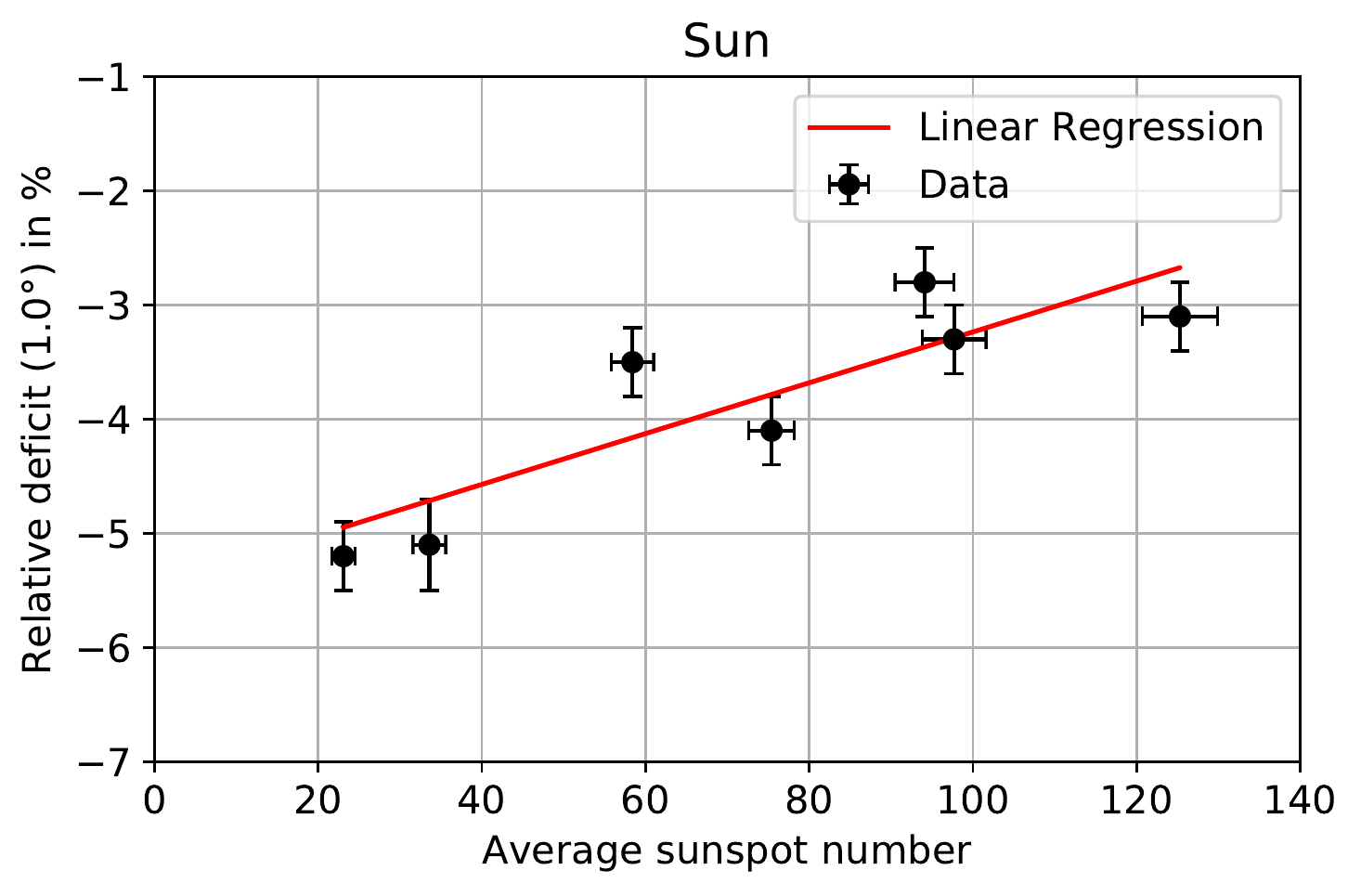}
    \caption{Correlation of measured relative deficit due to the Sun shadow and average sunspot number. A correlation of the two quantities is found to be likely.}
    \label{fig:RelDefSACorr}
\end{figure}

{}
In order to quantify the correlation between Sun shadow and solar activity, which is shown in Figure \ref{fig:RelDefSACorr}, two correlation tests are performed. 
The results of these tests are summarized in Table \ref{tab:sim_corr_test}.
While a Spearman's rank correlation test yields a correlation coefficient of 0.86 and a $p$-value of \SI{1.4}{\percent} for a correlation by chance, a Kendall-$\tau$ test yields a correlation coefficient of 0.71 and a $p$-value of \SI{3.0}{\percent}. 

We also quantify the deviation from a constant function by  fitting a linear function and a constant one. We calculate the difference in the $\chi^2$ and number of degrees of freedom, $\Delta \chi^{2}$ and $\Delta n_{\rm dof}$, respectively. Based on these values, the $p$-value is calculated from the cumulative distribution function of the appropriate $\chi^{2}$ distribution. While we find $\chi/n_{\rm dof} = 13.23/5$ for the linear model, the constant model yields $\chi/n_{\rm dof} = 53.67/6$. The difference in these two models therefore can be quantified to 
\begin{equation}
    \frac{\Delta \chi^{2}}{\Delta n_{\rm dof}}=\frac{40.44}{1}=40.44\,.
\end{equation}
This results in a $p$-value of $p=2.0\cdot 10^{-10}$, corresponding to a significance of $6.4\sigma$ that the linear fit is preferred over a constant one.
{}
\begin{table}[htbp]
\centering
\caption[Results: correlation of Sun shadow and solar activity]{Correlation coefficient and $p$-value of the two performed correlation tests.}
\label{tab:sim_corr_test} 
\begin{tabular}{@{}l|l|ll@{}}
\toprule
Correlation test & Correlation coefficient & $p$ in \si{\percent} \\
\midrule
Spearman's $\rho$ 	& 0.86 	& 1.4 \\
Kendall's $\tau$ 		&  0.71 	&	3.0 \\
\bottomrule
\end{tabular}
\end{table}
%

\subsection{\label{ssec:results_models}Comparison to solar magnetic field models}
\begin{figure}[htbp]
    \centering
    \includegraphics[width=\linewidth]{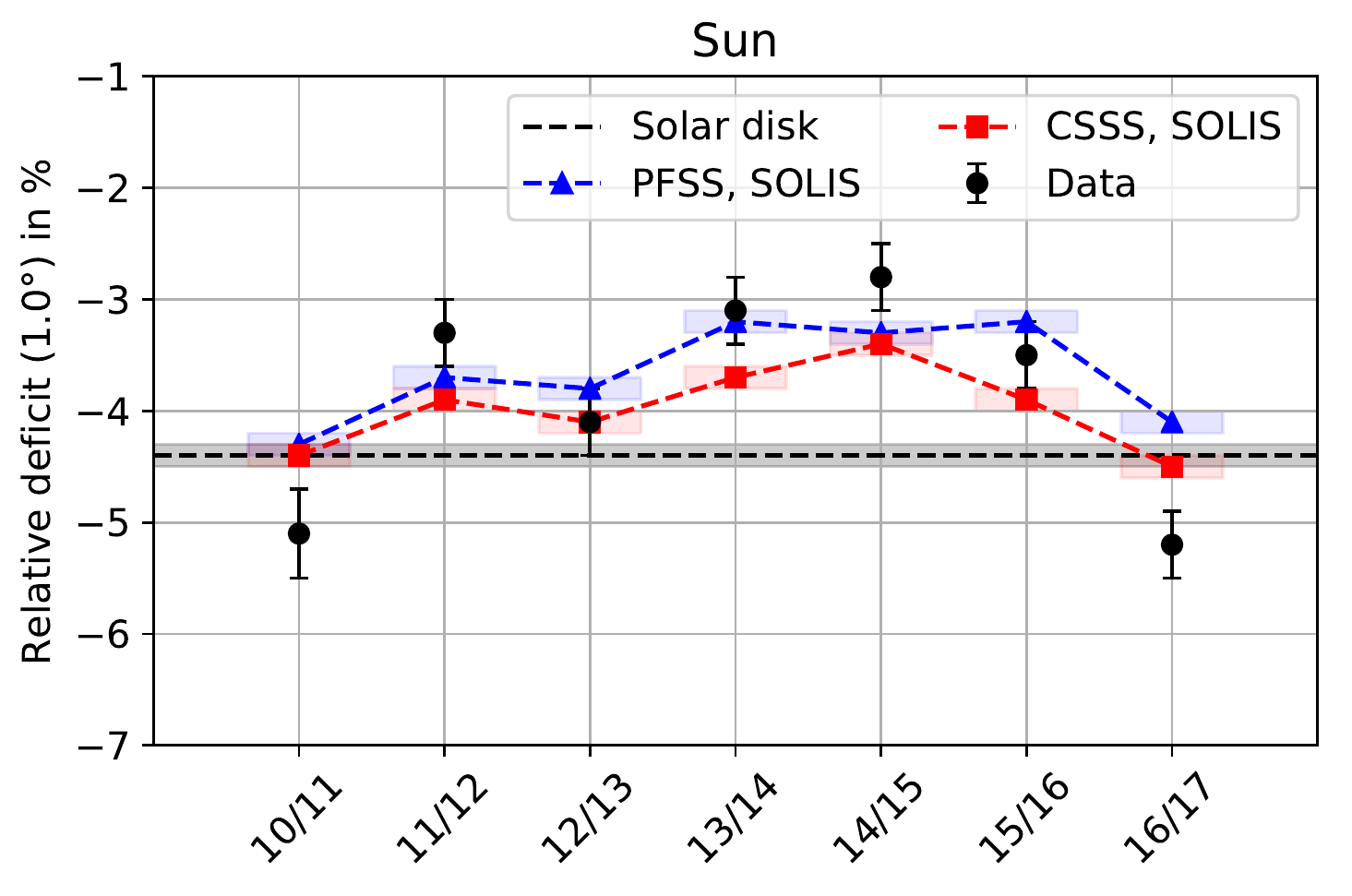}
    \caption{Comparison of measured relative deficit due to the Sun shadow and relative deficit expected from different models of the solar magnetic field.}
    \label{fig:SunModels}
\end{figure}

Finally, we modeled cosmic-ray propagation in the solar magnetic field to obtain predictions for the Sun shadow as expected from different coronal magnetic field models. 
Figure \ref{fig:SunModels} shows the results in terms of the observed relative deficit due to the Sun shadow and the expected relative deficit based on the PFSS and CSSS models in combination with the Parker spiral model introduced in Section \ref{ssec:analysis_sim_models}.

Both models reproduce the observed weakening of the shadow in times of high solar activity. 
The PFSS model predicts a more pronounced weakening of the shadow than the CSSS model in all years that are studied.
In 2010/2011 as well as in 2016/2017 the relative deficit observed in the data is slightly stronger than the prediction from both models, but also stronger than the expectation from the solar disk. 
{}
In addition to a $\chi^2$-test taking into account the statistical uncertainties of the data points, a modified $\chi^2$-test, which also takes into account the statistical uncertainties of the simulations and an estimate of the systematic uncertainty, is performed:
\begin{align}
\chi^2 = \sum_i \frac{(x_{\mathrm{data}}^i - x_{\mathrm{sim}}^i)^2}{\left({\sigma_{\mathrm{data}}^i}\right)^2 + \left({\sigma_{\mathrm{sim}}^i}\right)^2 + \sigma_{\mathrm{sys}}^2} \,.
\label{eq:sim_mod_chi2}
\end{align}
As in \cite{Tibet2013}, the systematic uncertainty is estimated from comparing the observed and expected relative deficit due to the Moon shadow and is found to amount to about \SI{0.3}{\percent}. 

While the standard $\chi^2$-test yields tensions between the data and the models on the order of 3 standard deviations, the modified test yields reasonable agreement with $p$-values of \SI{13}{\percent} and \SI{17}{\percent} for the PFSS and CSSS model, respectively.
All values are given in Table \ref{tab:sim_model_comparison}. These results can be compared to the findings of Tibet AS-$\gamma$ in \citep{Tibet2013}, who performed a similar study for the previous solar cycle (1996 -- 2009) at lower energies of $\sim10$~TeV. In \cite{Tibet2013}, it is discussed that the simulations with the CSSS model produce results that are consistent with the data. At these energies, the discrepancy with the PFSS model is larger ($p$-value of $4.9\cdot 10^{-5}$). In this paper, the CSSS model also fits somewhat better than the PFSS model, but these differences are not significant. One reason could be that the magnetic activity in solar cycle 24, which is investigated in this paper, is much less pronounced compared to solar cycle 23 which was investigated with Tibet AS-$\gamma$ data. In addition, effects should be amplified at lower energies as lower-energy particles have smaller gyro radii and therefore react stronger to the magnetic field. Finally, in this paper, we investigate seven years, while Tibet AS-$\gamma$ could make use of 14 years of data. 
\begin{table}[htbp]
\centering
\caption[Results: comparison with magnetic field models]{Reduced $\chi^2$, $p$-value, and significance of the comparison between the measured Sun shadows and the two different models.  Both models were provided with SOLIS magnetogram data \citep{SOLIS2020}. Values in parentheses are based on the modified $\chi^2$ given in equation \eqref{eq:sim_mod_chi2}.}
\label{tab:sim_model_comparison} 
\begin{tabular}{@{}l|l|l|l|l@{}}
\toprule
Magnetogram & Coronal model & $\chi^2 / n_{\mathrm{dof}}$ & $p$ & $S$ in $\sigma$ \\
\midrule
SOLIS & PFSS & 3.1 (1.6) & \num{0.0027} (\num{0.13}) & \num{3.0} (\num{1.5}) \\
SOLIS & CSSS & 2.9 (1.5) & \num{0.0052} (\num{0.17}) & \num{2.8} (\num{1.4})  \\
\bottomrule
\end{tabular}
\end{table}
%
\subsection{\label{ssec:results_energy}Energy dependence}
\begin{figure}[h!]
    \centering
    \includegraphics[width=\linewidth]{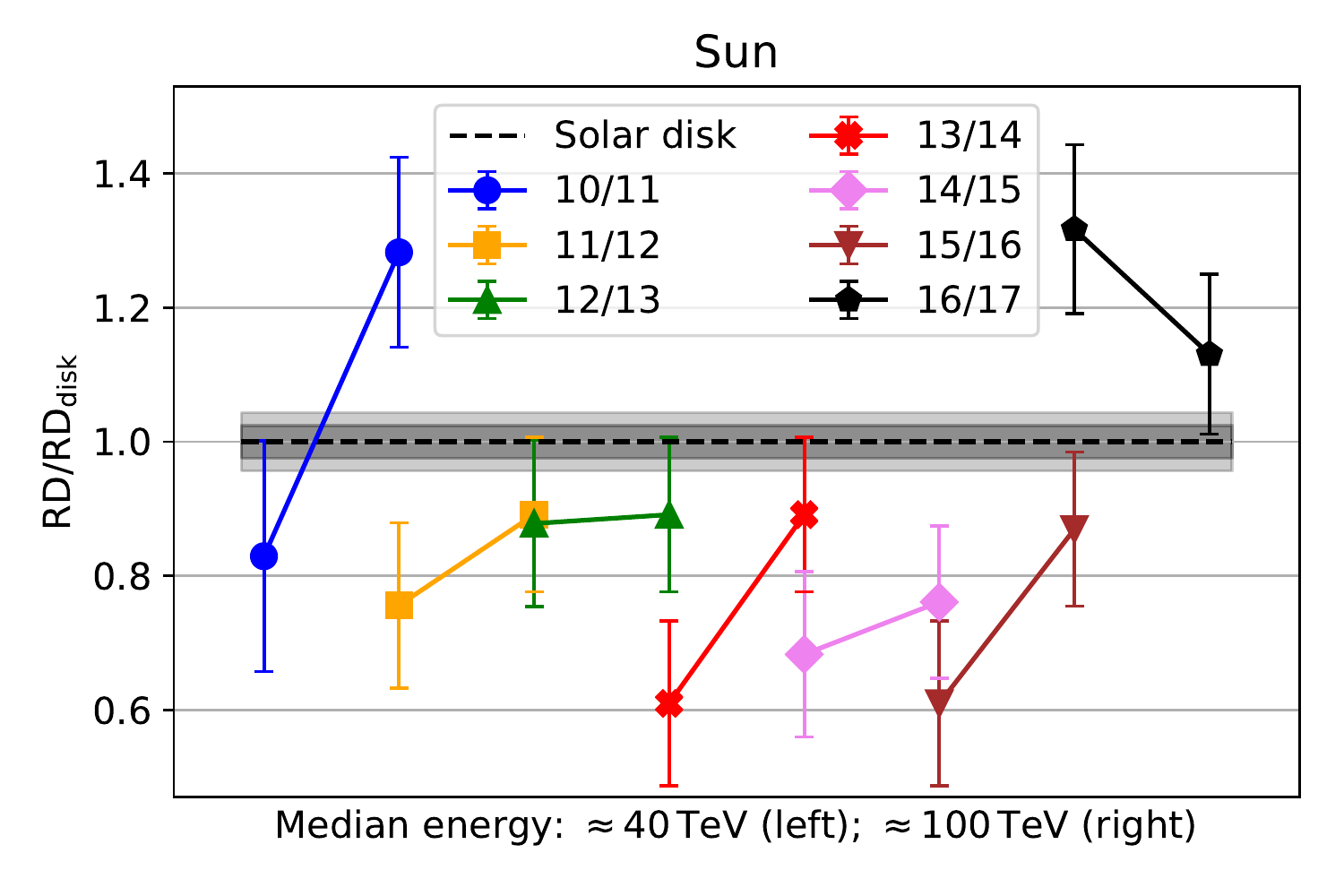}
    \caption{Relative deficit due to the Sun shadow normalized to the expectation from the solar disk as a function of the median energy of the sub-samples. For better visualization, pairs of data points for a specific season where shifted along the abscissa. The inner (outer) grey band indicates the statistical uncertainty of the disk-prediction for the low-energy (high-energy) sample.}
    \label{fig:EnergyDep}
\end{figure}

In this section we discuss the energy dependence of the Sun shadow. 
In Figure \ref{fig:EnergyDep} the relative deficit, normalized to the expectation from the solar disk, is shown for the low-energy (median energy: $\sim \SI{40}{\TeV}$) and high-energy (median energy: $\sim \SI{100}{\TeV}$) sub-samples during the observation period from November 2010 through February 2017.
Normalizing to the solar disk is necessary as the PSF is energy-dependent, causing the solar-disk shadow to be stronger for higher energies. 

For years with rather high solar activity (2011/2012--2015/2016) there is an indication for increasing shadow strength. 
For the two low-solar-activity years (2010/2011 and 2016/2017), no conclusions can be drawn. In general, these results are consistent with what is expected from theory \citep{beckertjus2020}, where it is shown that the shadow should increase in strength for years of high magnetic turbulence level and that it should be decreasing for low-activity years. However, to confirm such trends, a better energy resolution and a larger data set are necessary, therefore we refer to future work to investigate this question in more detail.
{}
%
\section{\label{sec:con}Conclusion}
In this paper, the time-dependent cosmic-ray Moon and Sun shadows were studied using seven years of IceCube data taken between May 2010 and May 2017. 
Both, Moon and Sun shadows are observed with high statistical significance in all seven years of data.
While the Moon shadow is described reasonably well by the lunar-disk model ($p=0.32$), the Sun shadow is statistically incompatible with geometrical shadowing only due to the solar disk ($7.3\sigma$).

We compared the temporal variation of the measured relative deficit of the Sun shadow to the change in the International Sunspot Number as a tracer for solar magnetic activity. 
We find the probability to observe the measured correlation by chance to be \SI{3.0}{\percent} (Kendall's $\tau$ test) or \SI{1.4}{\percent} (Spearman's rank test), respectively.
A linear relationship between shadow strength and solar activity is  preferred over a constant one with $6.4\sigma$.

We test two coronal magnetic field models, the PFSS and CSSS models, together with a Parker spiral beyond $2.5$ solar radii. 
Taking into account only statistical uncertainties, we find tensions between data and models on the order of $\sim 3\sigma$.
Including an estimate of the systematic uncertainty based on the observed Moon shadow, however, we compute reasonable $p$-values of \SI{13}{\percent} and \SI{17}{\percent} for the two models, respectively.

In times of high solar activity, the measured Sun shadow seems to increase with energy ($1.8\sigma$ indication).
In times of low solar activity, more data and an improved energy estimation will be necessary.

Future possibilities furthermore include testing different coronal magnetic field models like more general force-free or MHD models, studying the influence of CMEs on the Sun shadow, or implementing different models of the heliospheric magnetic field and/or the radial wind velocity profile. Further, in the future, energy-dependence can help to understand the strength of the field in particular when studying years of low solar activity. Here, an approximation by a Parker spiral is motivated by theory \citep{beckertjus2020,review2020}. A dipole-type field shows an increase of the shadow depth up to a certain maximum, which has a larger shadowing effect than the geometrical shadow. The predicted behavior of the Sun shadow during low solar activity with increasing energy is a monotonous increase of the shadow, converging toward the geometrical shadow, thus for all energies below convergence showing a shadow that is weaker than the geometrical one. Thus, for low activity years in particular, the observed peak energy can reveal the true normalization of the dipole, while the high activity years can help to disentangle the role of the small-scale component of the field. 
\begin{acknowledgements}
USA {\textendash} U.S. National Science Foundation-Office of Polar Programs,
U.S. National Science Foundation-Physics Division,
Wisconsin Alumni Research Foundation,
Center for High Throughput Computing (CHTC) at the University of Wisconsin-Madison,
Open Science Grid (OSG),
Extreme Science and Engineering Discovery Environment (XSEDE),
U.S. Department of Energy-National Energy Research Scientific Computing Center,
Particle astrophysics research computing center at the University of Maryland,
Institute for Cyber-Enabled Research at Michigan State University,
and Astroparticle physics computational facility at Marquette University;
Belgium {\textendash} Funds for Scientific Research (FRS-FNRS and FWO),
FWO Odysseus and Big Science programmes,
and Belgian Federal Science Policy Office (Belspo);
Germany {\textendash} Bundesministerium f{\"u}r Bildung und Forschung (BMBF),
Deutsche Forschungsgemeinschaft (DFG),
Helmholtz Alliance for Astroparticle Physics (HAP),
Initiative and Networking Fund of the Helmholtz Association,
Deutsches Elektronen Synchrotron (DESY),
and High Performance Computing cluster of the RWTH Aachen;
Sweden {\textendash} Swedish Research Council,
Swedish Polar Research Secretariat,
Swedish National Infrastructure for Computing (SNIC),
and Knut and Alice Wallenberg Foundation;
Australia {\textendash} Australian Research Council;
Canada {\textendash} Natural Sciences and Engineering Research Council of Canada,
Calcul Qu{\'e}bec, Compute Ontario, Canada Foundation for Innovation, WestGrid, and Compute Canada;
Denmark {\textendash} Villum Fonden, Danish National Research Foundation (DNRF), Carlsberg Foundation;
New Zealand {\textendash} Marsden Fund;
Japan {\textendash} Japan Society for Promotion of Science (JSPS)
and Institute for Global Prominent Research (IGPR) of Chiba University;
Korea {\textendash} National Research Foundation of Korea (NRF);
Switzerland {\textendash} Swiss National Science Foundation (SNSF);
United Kingdom {\textendash} Department of Physics, University of Oxford.
\end{acknowledgements}
\providecommand{\noopsort}[1]{}\providecommand{\singleletter}[1]{#1}%


\end{document}